\documentclass[twocolumn]{aastex631}

\graphicspath{{./}{figures/}}

\received{\today}
\submitjournal{ApJs}

\shortauthors{Coppi et al.}

\usepackage{amsmath}
\usepackage[normalem]{ulem}
\let\tablenum\relax
\usepackage{siunitx}
\usepackage{xurl}
\DeclareSIUnit{\bel}{B}
\DeclareSIUnit{\dB}{\deci\bel}
\DeclareSIUnit{\dBm}{\deci\bel\meter}

\begin{document}

\title{PROTOCALC, A W-band polarized calibrator for CMB Telescopes: application to Simons Observatory and CLASS}

\correspondingauthor{Gabriele Coppi}
\email{gabriele.coppi@unimib.it}

\author[0000-0002-6362-6524]{Gabriele Coppi}
\affiliation{Department of Physics, University of Milano-Bicocca, Piazza della Scienza 3, 20126, Milano, Italy}
\affiliation{Istituto Nazionale di Fisica Nucleare, INFN, Sezione Milano-Bicocca, Piazza della Scienza 3, 20126, Milano, Italy}

\author[0009-0006-7382-1434]{Nadia Dachlythra}
\affiliation{Department of Physics, University of Milano-Bicocca, Piazza della Scienza 3, 20126, Milano, Italy}

\author[0000-0002-8307-5088]{Federico Nati}
\affiliation{Department of Physics, University of Milano-Bicocca, Piazza della Scienza 3, 20126, Milano, Italy}
\affiliation{Istituto Nazionale di Fisica Nucleare, INFN, Sezione Milano-Bicocca, Piazza della Scienza 3, 20126, Milano, Italy}

\author[0000-0003-3892-1860]{Rolando D\"{u}nner-Planella}
\affiliation{Instituto de Astrof\'{i}sica and Centro de Astro-Ingenier\'{i}a, Facultad de F\'{i}sica, Pontificia Universidad Cat\'{o}lica de Chile, Av. Vicu\~{n}a Mackenna 4860, 7820436, Macul, Santiago, Chile}

\author[0000-0002-5736-5524]{Alexandre E. Adler}
\affiliation{Department of Physics, University of California, Berkeley, 366 LeConte Hall Berkeley, CA
94720, USA}
\affiliation{Physics Division, Lawrence Berkeley National Laboratory, 1 Cyclotron Road, Berkeley, CA
94720, USA}

\author[0000-0002-1419-0031]{Josquin Errard}
\affiliation{Universit\'e Paris Cit\'e, CNRS, Astroparticule et Cosmologie, F-75013 Paris, France}

\author[0000-0001-7225-6679]{Nicholas Galitzki}
\affiliation{Department of Physics, University of Texas at Austin, Austin, TX, 78712, USA}
\affiliation{Weinberg Institute for Theoretical Physics, Texas Center for Cosmology and Astroparticle Physics, Austin, TX 78712, USA}

\author[0000-0002-4820-1122]{Yunyang Li}
\affiliation{Kavli Institute for Cosmological Physics, University of Chicago, 5640 South Ellis Avenue, Chicago, IL 60637, USA}

\author[0000-0002-4436-4215]{Matthew A. Petroff}
\affiliation{Center for Astrophysics, Harvard \& Smithsonian, Cambridge, MA 02138, USA}

\author[0000-0001-9221-7802]{Sara M. Simon}
\affiliation{Fermi National Accelerator Laboratory, Batavia, IL, USA}

\author[0009-0001-6108-9518]{Ema Tsang King Sang}
\affiliation{Universit\'e Paris Cit\'e, CNRS, Astroparticule et Cosmologie, F-75013 Paris, France}

\author[0009-0004-4775-9935]{Amalia Villarrubia Aguilar}
\affiliation{Universit\'e Paris Cit\'e, CNRS, Astroparticule et Cosmologie, F-75013 Paris, France}

\author[0000-0002-7567-4451]{Edward J. Wollack}
\affiliation{NASA Goddard Space Flight Center, 8800 Greenbelt Road, Greenbelt, MD 20771, USA}.

\author[0000-0002-4495-571X]{Mario Zannoni}
\affiliation{Department of Physics, University of Milano-Bicocca, Piazza della Scienza 3, 20126, Milano, Italy}
\affiliation{Istituto Nazionale di Fisica Nucleare, INFN, Sezione Milano-Bicocca, Piazza della Scienza 3, 20126, Milano, Italy}

\begin{abstract}

Current- and next-generation Cosmic Microwave Background (CMB) experiments will measure polarization anisotropies with unprecedented sensitivities. The need for high precision in these measurements underscores the importance of gaining a comprehensive understanding of instrument properties, with a particular emphasis on the study of the beam properties and, in particular, their polarization characteristics, and the measurement of the polarization angle. In this context, a major challenge lies in the scarcity of millimeter polarized astrophysical sources with sufficient brightness and calibration knowledge to meet the stringent accuracy requirements of future CMB missions. This led to the development of a drone-borne calibration source designed for frequency band centered on approximately \SI{90}{\giga\hertz} band, matching a commonly used channel in ground based CMB measurements. The PROTOtype CALibrator for Cosmology, PROTOCALC, has undergone thorough in-lab testing, and its properties have been subsequently modeled through simulation software integrated into the standard Simons Observatory (SO) analysis pipeline. Moreover, the PROTOCALC system has been tested in the field, having been deployed twice on calibration campaigns with CMB telescopes in the Atacama desert. The data collected constrain the roll angle of the source with a statistical accuracy of \SI{0.045}{\degree}.

\end{abstract}


\section{Introduction} \label{sec:intro}

Cosmic Microwave Background (CMB) observations have been foundational to our understanding of the evolution of the Universe and its energy content. Measurements of the small-scale CMB temperature and polarization fields allow us to constrain the cosmological parameters, measure the effective number of relativistic species, and set limits to primordial non-Gaussianity \citep{Planck_IX_2020, Planck_VI_2020}. Furthermore, the large-scale CMB power spectrum retains an almost scale-invariant form and can thus be employed to study the early Universe and initial conditions \citep{Tristram_2021}. A particularly interesting aspect which refers to the CMB polarization is that it can be decomposed into a strong, curl-free, and even-parity component, E-modes, and a fainter odd-parity and divergence-free component, B-modes \citep{Kosowsky_1996}. A key difference between the two components is that, while $E$-modes can be produced both by scalar and tensor perturbations of the space-time metric, the primordial $B$-modes the CMB community is seeking can only be sourced by tensor perturbations. Although the CMB community has obtained precise measurements of the $E$-mode spectra \citep{Louis_2017, Planck_V_2020, Dutcher_2021}, a detection of primordial $B$-modes has not yet been achieved. Such detection would provide a significant probe of cosmic inflation \citep{Kamionkowski_1997, Seljak_1997}, the current leading theory for initial perturbations, according to which an initially small patch of the Universe underwent abrupt, exponential expansion at very early times. According to most inflationary models, an inevitable consequence in this scenario would be the creation of ripples in space-time that are expressed through tensor perturbations of the metric and can be thus probed via the CMB $B$-modes \citep{Kamionkowski_2016}. To quantify the inflationary gravitational wave signal we can detect from the CMB maps, we employ the so-called tensor-to-scalar ratio, $r$, i.e. the ratio between the amplitude of tensor and the scalar perturbations. 
The present upper-limit on the scalar-to-tensor ratio is, $r < 0.036$, at a $95\%$ confidence level (C.L.) and associated uncertainty of $\sigma(r) = 0.009$ \citep{bicep_keck_r_constraints_2021}. Current-generation experiments like the Simons Observatory (SO) Small Aperture Telescopes (SATs) aim at constraining the tensor-to-scalar ratio with a statistical uncertainty of $\sigma(r)=0.003$ or better \citep{SO_goals_forecasts}, while next-generation telescopes like LiteBIRD and CMB-S4 will attempt a $r<0.003$ detection with a statistical uncertainty of $\sigma(r)<0.001$ \citep{Litebird_mission_2020, s4collaboration2020cmbs4}.

Measuring $r$ to that level of precision demands exceptionally high accuracy in characterizing all potential systematic effects, achieved through a thorough understanding of the instrument's performance. In particular, great caution must be taken when calibrating the spatial response of the optical system - commonly referred to as the instrument's \textit{Point-Spread-Function} (PSF) or \textit{beam} - and the absolute orientation of each polarization-sensitive detector in the telescopes’ focal planes, commonly referred to as the absolute polarization angle. Beam systematics such as ellipticity, increased power at large angles away from the beam center (sidelobes), or cross-polar sensitivity may cause the strong temperature signal to leak into the fainter polarization data or lead to mixing between the two polarization components \citep{Hu_2003, Miller_beam_asymmetry, bicep2_systematics, beamconv_2018, Lungu_2022}. These are commonly modeled relying on planet observations. However, planets are not always sufficiently bright/available to reach the required calibration accuracies and, being effectively unpolarized, can not be used to calibrate the polarization angle. A miscalibrated polarization angle can assign part of the $E$-mode signal to the CMB $B$-mode component, affecting the $r$-tensor constraint and producing a spurious signal that could be mistaken for cosmic birefringence. Birefringence describes signature of parity-violating physics on the CMB and can serve as a probe of physical processes beyond the Standard Model of cosmology \citep{Carrol1990, Carrol1998, cosmic_birefringence}. For CMB experiments, one can correct the polarization angle by assuming that any non-zero CMB $EB$ spectra are the result of an angle calibration error \citep{Keating_2013, Krachmalnicoff_inflight_polang}. This error can be either simply corrected in the analysis or further studied as the combination of a true instrumental miscalibration and a rotation due to cosmic birefringence \citep{Jost2023, delahoz2022}. However, in order to meaningfully separate these two contributions, the detectors' polarization angles of the experiment in question must already be calibrated with exceptional precision \citep{Navaroli_2018, Minami_2019, Murphy_2024}. For the reasons outlined above, it is clear that employing a strongly emitting, fully polarized millimeter-wave source is crucial for achieving the highest calibration accuracy in current and next-generation CMB instruments.

In practice, the population of astrophysical sources with brightness well matched to the calibration needs of CMB experiments is limited and is further reduced upon considering polarization. Consideration of the subset sources that are also polarized further reduces the number of potential calibration candidates. Planets are the most common candidates for calibrating CMB experiments \citep{calibration_wmap_2011, Hasselfield_2013, planck_vii_2016, Lungu_2022,
Dachlythra_2024}, but they are often subject to availability issues when scanning from the ground and are very faintly polarized (maximum polarization fraction corresponds to Uranus and is less than $3.6\%$ according to \cite{Planck_LII_2017}). The most promising natural candidate for polarization is Tau A (the Crab Nebula) and allows for the determination of the average polarization angle with accuracy $\Delta\psi\simeq 0.33\si{\SIUnitSymbolDegree}$ \citep{Aumont_2021}. This will not be sufficient for the $\leq$ \SI{0.2}{\degree} polarization angle accuracy required at \SI{90}{} and \SI{150}{\giga\hertz} by modern experiments like SO \citep{Abitbol_2021}, aiming to constraints the tensor-to-scalar ratio with biases smaller than $\Delta r \simeq 10^{-3}$. It is also reasonable to expect that this requirement will become even more stringent for upcoming ground-based missions such as CMB-S4 \citep{s4collaboration2020cmbs4}. Instead, for cosmic birefringence measurements, the requirements are even more stringent \citep{Bicep2025}. Finally, it is important to highlight that experiments like SO presents additional challenges in polarization angle calibration due to the frequency-dependent behavior of the Half-Wave Plate \citep{Verges2021, Vielva2022}. This need has motivated the development of artificial sources. As the calibration source needs to be placed in the telescope far-field, a variety of methods have been employed, including mounting an artificial source on a tall structure \citep{BK2019} or creating a space-based calibrator \citep{Johnson2015}. 

In this work, we introduce an aerial drone-based approach, PROTOCALC, along with its successor, POLOCALC \citep{Nati2017}, developed to calibrate various CMB telescopes, with a particular focus on experiments observing from the Atacama Desert in Chile. The paper is organized as follows: Section \ref{sec:protocalc} discusses the capabilities and design of the drone-based source. Section \ref{sec:laboratory} then describes the in-lab characterization of the source, while Section \ref{sec:simulation} quantifies its expected performance on the field via simulations. Finally, Section \ref{sec:inflight} presents initial results from a test flight over the observation site in the Atacama Desert.

\section{PROTOCALC} \label{sec:protocalc}

As already established in the Introduction, high-accuracy calibration of current and future CMB missions requires a source that will enable an extensive study of the beam properties and can produce a polarized signal that is sufficiently strong to calibrate the telescope's polarization angle with an accuracy better than $\SI{0.2}{\degree}$. In this scope, we developed a Drone-based source called PROTOCALC. This concept was also developed for the \SI{150}{\giga\hertz} band \citep{Dunner2020}. This source is designed to calibrate the instrument's polarized response in the W-band, so between $75$ and \SI{115}{\giga\hertz} covering the W band of CLASS \citep{Dahal2022} and one of the Middle Frequency (MF) bands of Simons Observatory \citep{SO_goals_forecasts}, and is expected to achieve a polarization angle accuracy of $0.1\si{\SIUnitSymbolDegree}$. 

\subsection{UAV Platform} \label{sec:drone}
\label{UAV_platform}
\begin{figure}
    \centering
    \includegraphics[width=\linewidth]{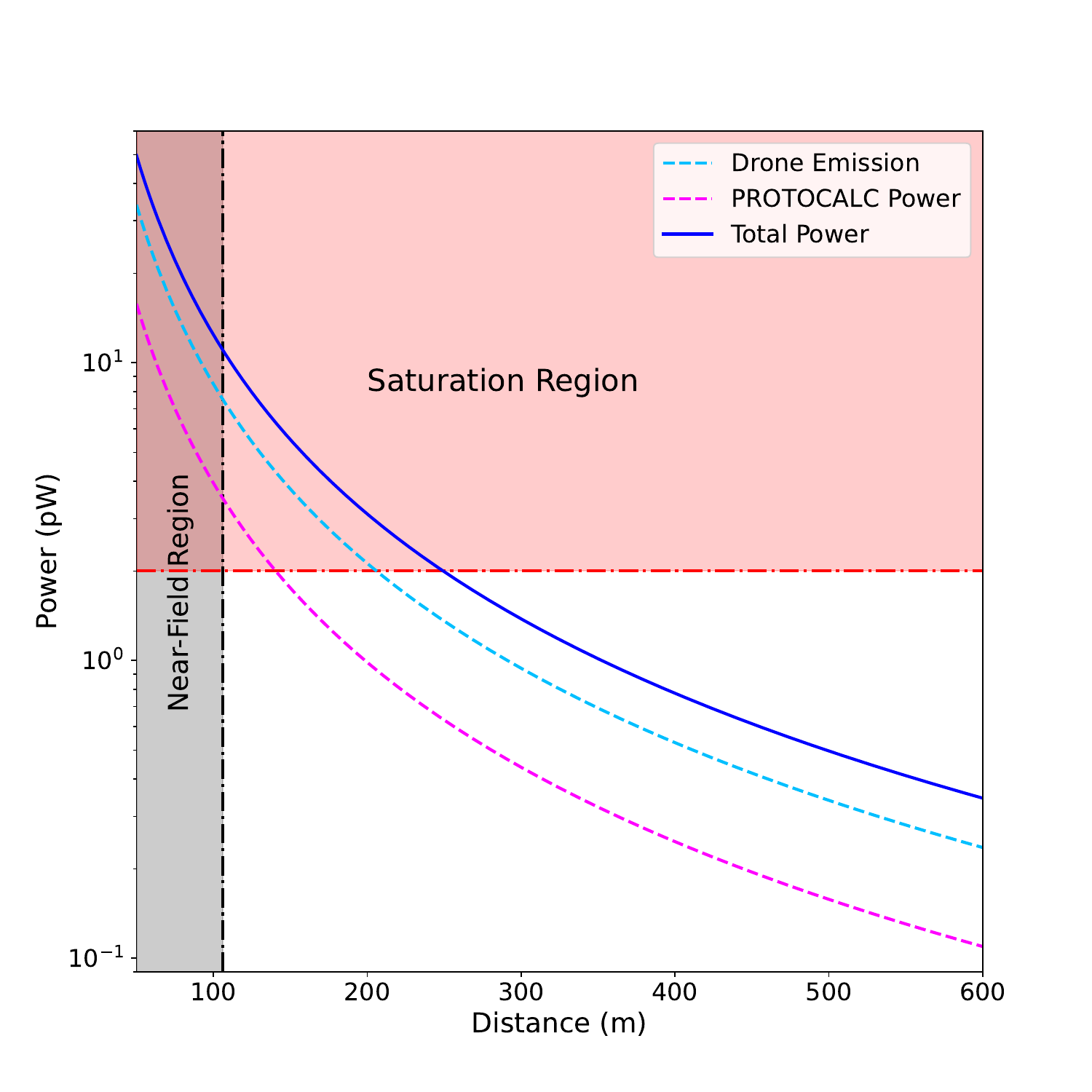}
    \caption{Emission of the drone modeled as a blackbody in the \SI{90}{\giga\hertz} band with a bandwidth of $30\%$, received by a detector in a small aperture-type (like SO-SAT or CLASS) telescope with \SI{50}{\centi\meter} aperture. We included also the power emitted by the PROTOCALC source considering a diffraction-limited single mode telescope. The vertical dotted line represents the far-field distance, while the horizontal represents the \SI{2}{\pico\watt} power which is a conservative saturation power for TES detectors.}
    \label{fig:emission}
\end{figure}

To keep the setup as simple as possible, we decided to use a commercially available drone as a platform, the DJI Matrice 600 Pro. This Unmanned Aerial Vehicle (UAV) has a flight time of approximately \SI{20}{\minute} when at sea level and carrying to a payload weighing \SI{5}{\kilo\gram}. The flight time is reduced by $50\%$ when the drone is flown at high altitude, above \SI{5000}{\meter}. This measurement strategy hinges upon the stability and knowledge of the source polarization angle and a DJI gimbal (RONIN MX\footnote{\url{www.dji.com/ronin-mx}}) is employed to stabilize the payload in flight.
To assess the feasibility of the drone for use with the CLASS telescopes and SO-SATs \citep{Galitzki2024}, we calculate the thermal emission from the drone itself. The thermal emission must be low enough to prevent saturation of the telescope detectors while also allowing sufficient overhead to operate the calibration source. For the SO-SATs, the saturation power in the Mid-Frequency bands (MF), is \SI{5.2}{\pico\watt} and \SI{7.3}{\pico\watt} for two $25 \%$-width frequency bands centered around \SI{93}{} and \SI{145}{\giga\hertz}, respectively \citep{McCarrick2021}. Instead, CLASS presents higher saturation power at \SI{90}{} and \SI{150}{\giga\hertz}, with values of \SI{18.4}{\pico\watt} and \SI{35.0}{\pico\watt}, respectively \citep{Dahal2022}. As it is evident from Figure \ref{fig:emission}, the total power seen by any detector is less than a conservative value of \SI{2}{\pico\watt} at $\simeq$ \SI{250}{\meter} distance for the \SI{90}{\giga\hertz} SO frequency.

\begin{figure*}[t!]
    \centering
    \includegraphics[width=\textwidth, trim={2cm 0 2cm 0}]{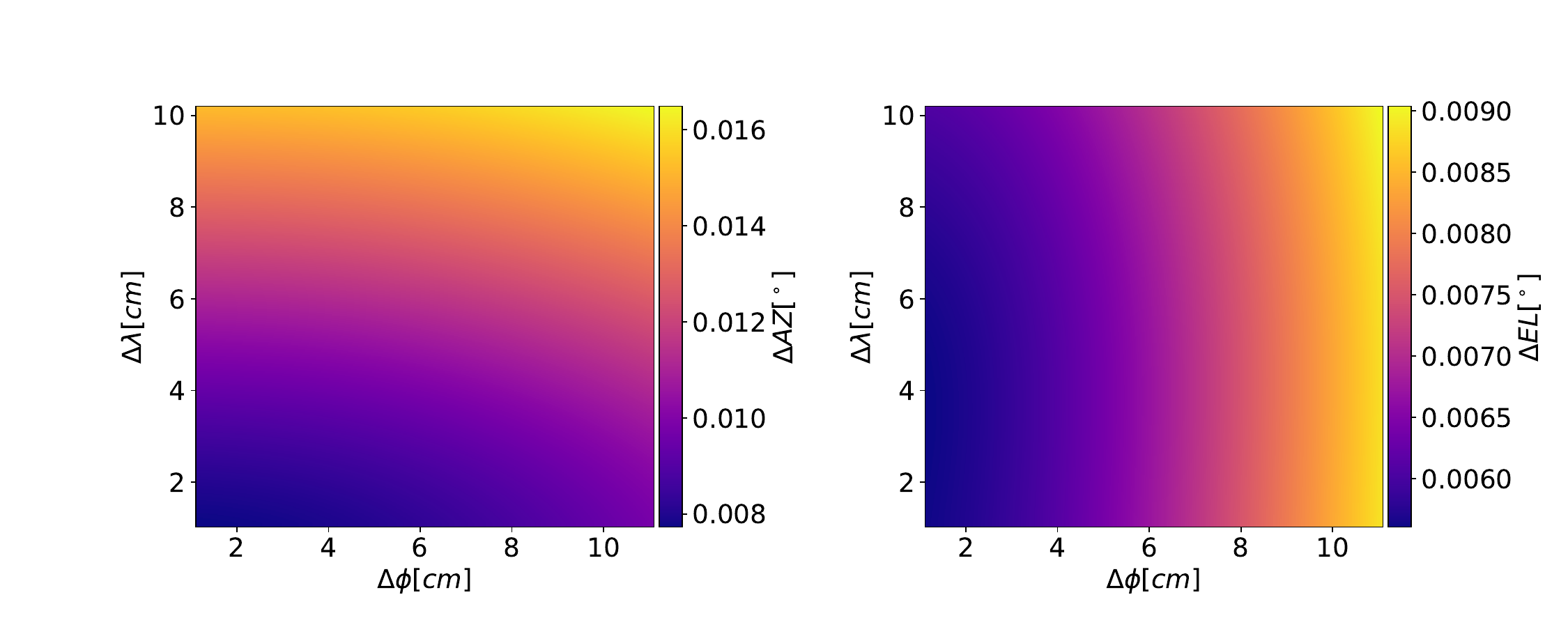}
    \caption{Error in azimuth and elevation coordinates of the drone at a distance of $\SI{500}{\meter}$ and $\SI{45}{\degree}$ elevation. The typical RTK uncertainty at the Atacama site is $\SI{6}{\centi\meter}$ on the vertical position. The plot shows a worst-case scenario of a constant $\SI{10}{\centi\meter}$ error.}
    \label{fig:azimuth}
\end{figure*}

Other than the emission of the UAV, it is also important that the platform position's is known well enough that the error in the telescope's pointing is negligible compared to required accuracy for the polarization angle. The position of the drone is provided by the on-board GPS using Real-Time Kinematics (RTK) techniques \citep{Manandhar1999}. RTK is a powerful position technique that is able to achieve an horizontal accuracy of down to \SI{1}{\centi\meter} and a vertical accuracy of down to \SI{2}{\centi\meter} with respect to a given base station. As shown in Figure \ref{fig:azimuth}, for a conservative accuracy of an RTK system (\SI{4}{\centi\meter} horizontal and \SI{10}{\centi\meter} vertical), the error in azimuth and elevation is lower than \SI{0.01}{\SIUnitSymbolDegree}, where we have considered the drone flying at an elevation of \SI{45}{\SIUnitSymbolDegree} and an azimuth of \SI{180}{\SIUnitSymbolDegree} with respect to the telescope. Additional information on the attitude determination will be presented in section \ref{subsec:attitude}.

\subsection{Mechanical Design and Power System} \label{subsec:design}

\begin{figure}[h!]
    \centering
    \includegraphics[width=\linewidth]{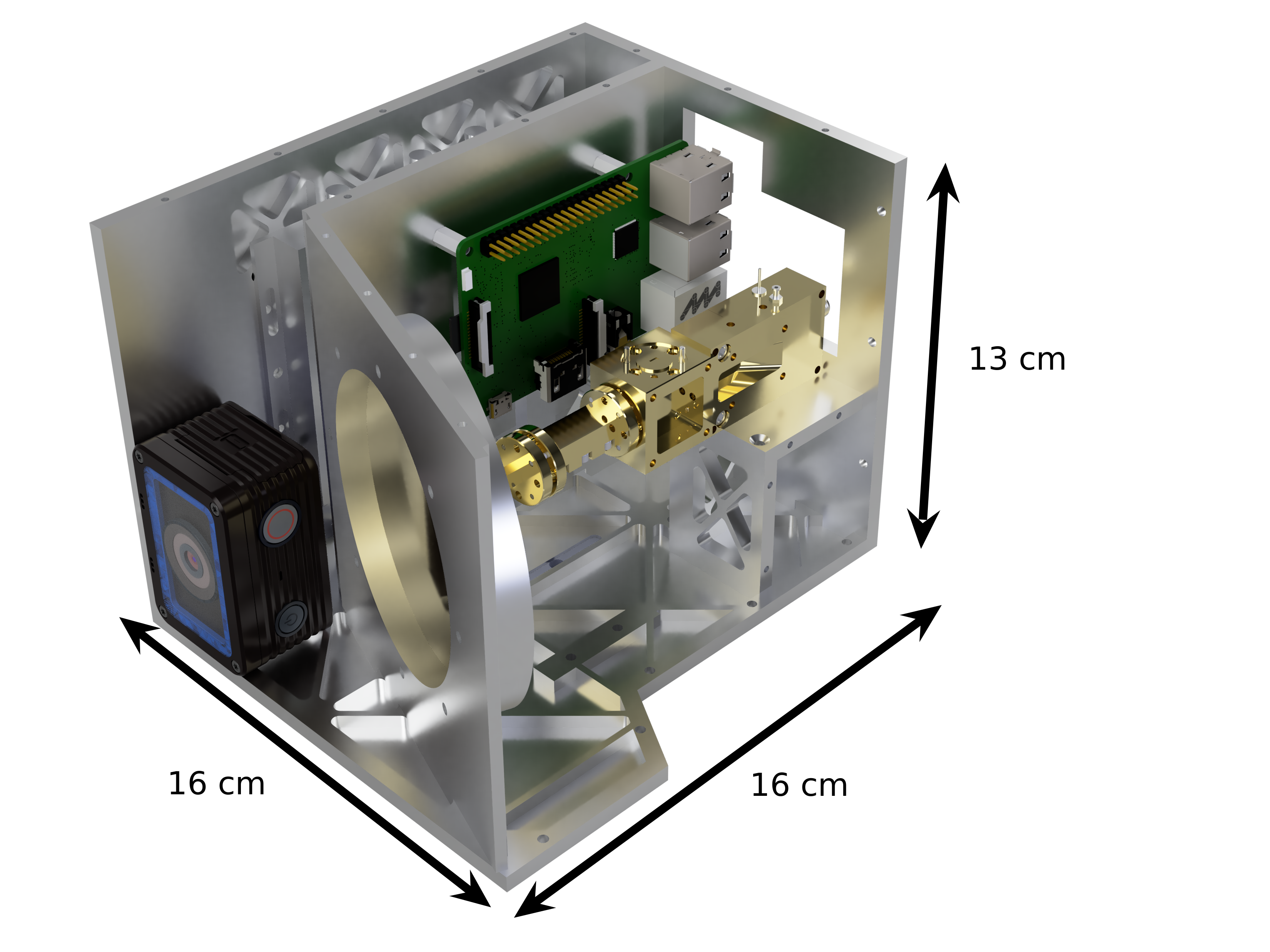}
    \caption{Render of the payload with all the components.}
    \label{fig:render}
\end{figure}

The achievable drone platform lift and gimbal mount volume define the allowable payload allocations for the polarization source module. According to specification, the gimbal weighs \SI{2.77}{\kilo\gram} while the maximum payload weight for the UAV is \SI{5.5}{\kilo\gram} at take-off\footnote{This value can be increased to \SI{6}{\kilo\gram} using lighter batteries with lower capacity}. Given the weight of the gimbal and the take-off limit of the UAV, the maximum allowed weight for the payload is $\SI{2}{\kilo\gram}$. The total available volume for the payload can be represented by a \qtyproduct[product-units=single]{16 x 16 x 13}{\centi\metre} box. 
The weight and size constraints are not the only ones that we need to consider while designing the source. Indeed, we also need to consider environmental constraints. In particular, PROTOCALC is designed to work at an altitude of \SI{5200}{\meter} where the pressure is roughly equal to \SI{500}{\milli\bar}. In a similar environment, passive convective cooling by air provides insufficient heat removal to maintain a viable temperature for the source active components. For simplicity, heat dissipated by active elements is conductively transferred through the payload mechanical bus and removed from the system via radiative cooling. Conduction is achieved through the use of high conductivity material for the payload, as Aluminum-6061.

In order to keep the weight of the payload minimized, we optimized the structure by removing weight creating a honeycomb structure that would maintain the structural rigidity of the full assembly. Additionally, we designed multiple components to enable easy access to the internal components in the field. This design also enabled more convenient testing in the lab in configurations where the setup was not fully assembled. The polarizing grid that defines the polarization axis is mounted at a $20\si{\SIUnitSymbolDegree}$ angle, with respect to the source aperture. Tilting the grid with respect to the antenna of a small angle ($< 25\si{\SIUnitSymbolDegree}$), prevents the formation of standing waves. We decided on the chosen value to keep the source as compact as possible. Given that the polarized signal coming from the antenna and the polarization axis of the grid are parallels, the output signal will not be affected. Small patches of Eccosorb HR-10\footnote{\url{https://www.laird.com/products/absorbers/microwave-absorbing-foams/single-layer-foams/eccosorb-hr}} absorber are attached to the interior of the source housing which realized from high reflectivity aluminum. This configuration prevents the formation of cavity resonances between the source and the polarizer and terminates reflected source power within the housing. The camera that is used for photogrammetry (see \ref{subsec:attitude}) is located outside of the cavity to allow the maximum available field of view without obstruction. One of the key components for accurate measurements of the drone-platform roll angle, which is used as a proxy for the source polarization angle in flight, is the correct alignment between the photogrammetry camera and grid polarizer as will be discussed later in section \ref{subsec:lab_opt_att}. The structural Al-6061 components are machined with a CNC achieving a manufacturing accuracy of $\simeq$ $\SI{10}{\micro\meter}$ and is subsequently aligned in the lab as explained in  \ref{subsec:lab_opt_att}.
The final dimensions and weight of the payload are within the envelope given by the gimbal corresponding to a volume of \qtyproduct[product-units=single]{16 x 13 x 13}{\cm} and weight equal to $\SI{1.75}{\kilo\gram}$. All the components on board are powered by the gimbal which provides a DC output of $\SI{13}{\volt}$ and $\SI{3}{\ampere}$ for a total available power of $\SI{39}{\watt}$. The total power consumption of the payload is around \SI{14}{\watt} and it is dominated by the frequency synthesizer (\SI{6}{\watt}), the frequency multiplier (\SI{4.5}{\watt}) and the Raspberry Pi (\SI{3}{\watt}). To power different components, we developed a custom board that takes in input the DC signal coming from the gimbal and converts it to required input voltage for each component. The gimbal's dynamic power consumption is $\SI{9.6}{\watt}$ and carries a battery of $\SI{22.75}{\watt\hour}$. As a result, the full system of gimbal and payload can be powered for almost one hour which is longer than the maximum flight time with \SI{5.5}{\kilo\gram} payload of \SI{16}{\minute} for the chosen UAV.

\begin{figure}
    \centering
    \includegraphics[width=\linewidth]{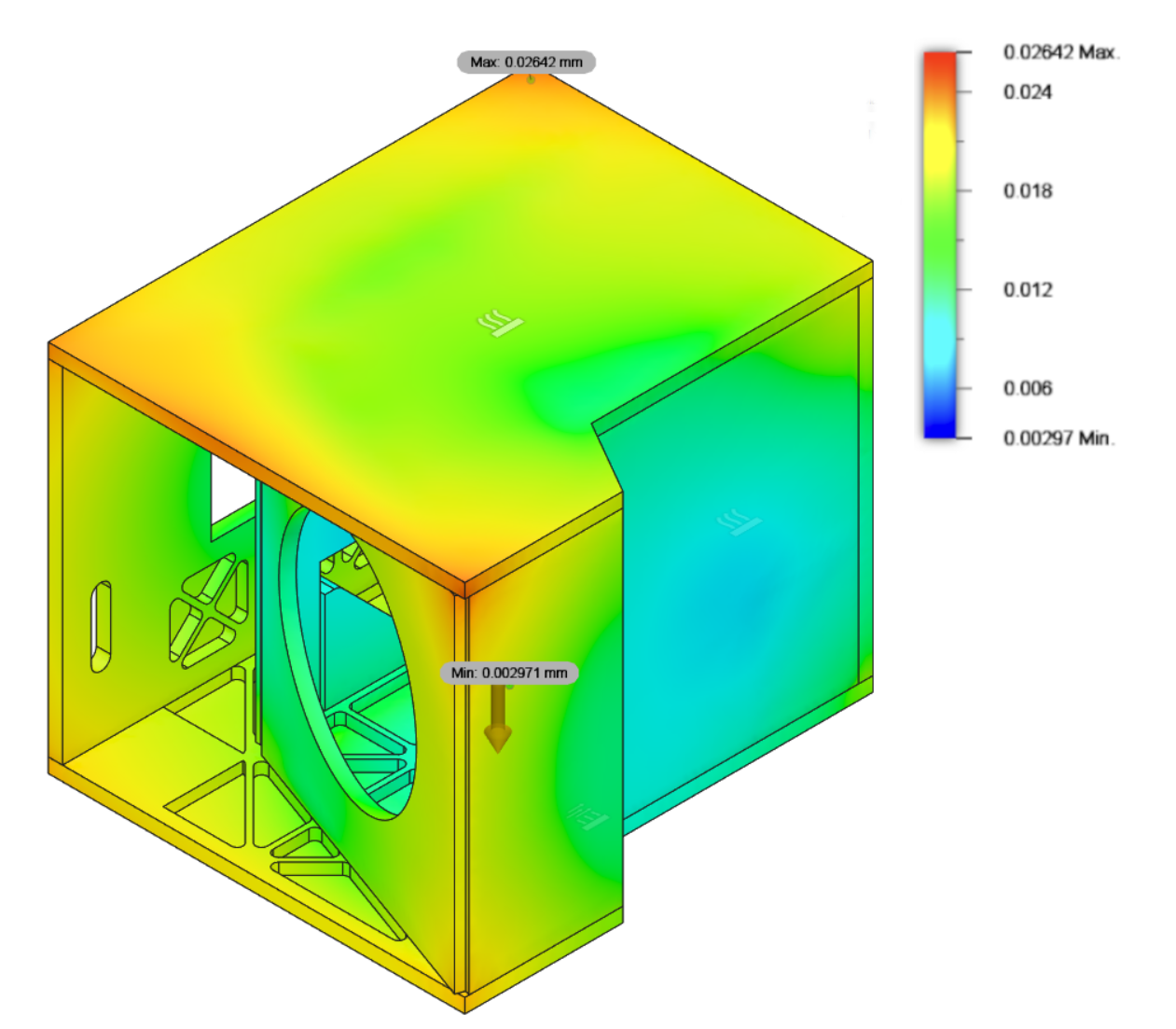}
    \caption{Displacement computed using a thermo-mechanical simulation on the payload.}
    \label{fig:displacement}
\end{figure}

\subsection{Mechanical and Thermal Simulations}\label{subsec:mec_simulations}

PROTOCALC is designed to calibrate telescopes situated at different observing sites; therefore, it is crucial to verify that the developed setup can be deployed in challenging environments such as the site at Cerro Toco in the Atacama Desert, where several CMB telescopes are located. Thermal simulations are used to verify the following design considerations: (1) all the components work within temperature specifications and (2) the differential contraction of the materials do not result in the misalignment of the optical components more than the required accuracy on polarization angle. To perform this assessment, we performed mechanical and thermal simulations using the Autodesk Fusion360 
software\footnote{\url{https://www.autodesk.com/products/fusion-360/}}.
For PROTOCALC, we setup the simulation considering a scenario where the only mechanical load is gravity and the mechanical constraints is the contact surface between the payload and the gimbal. Regarding the thermal loads, we included all the sources of heat introduced in subsection \ref{subsec:design} and we include the radiative cooling. Due to limitations on the drone operating temperatures, the calibrator will be deployed when the temperature is around \SI{0}{\celsius} or higher. We can, therefore, set this value as our environment temperature in the simulation. The thermal, stress-free temperature is considered to be \SI{20}{\celsius} which matches the value of the typical laboratory temperature where the payload is assembled and characterized. The results of the simulation are presented Figure \ref{fig:displacement}. The relative displacement between the camera and the holes from mounting the polarization grid is negligible ($\le \SI{10}{\micro\meter}$). The final temperature is predicted to be between \SI{6}{} and \SI{12}{\celsius}, which is well within the operating range for all the components used for the project. These results demonstrate that the choice of aluminum is justified. Indeed, the high conductivity of aluminum leads the payload temperature to not vary significantly, and this is reflected in small relative displacement between the components.

\subsection{RF Configuration}\label{subsec:RFConfiguration}

\begin{figure*}
    \centering
    \includegraphics[width=\textwidth, trim={2cm 0 2cm 0}]{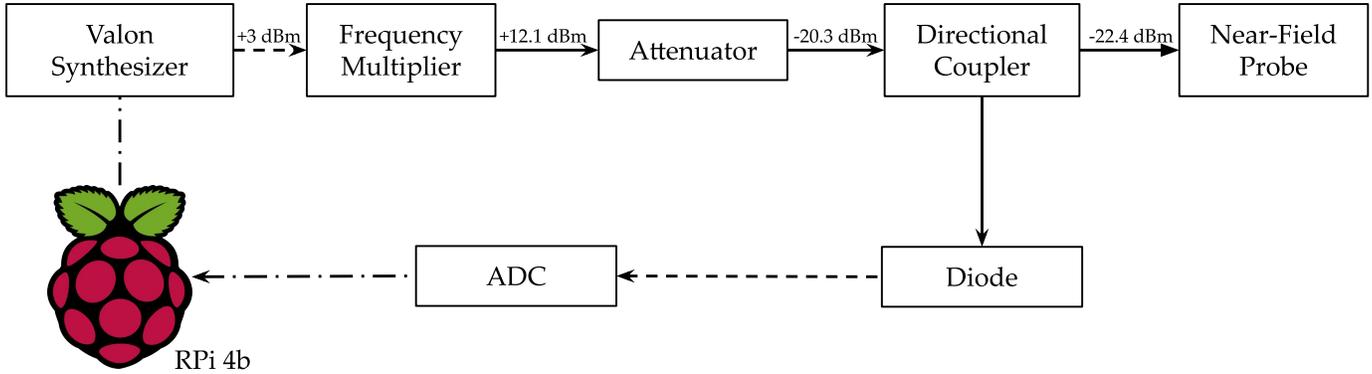}
    \caption{RF schematic output of PROTOCALC. We present also the output power from each components as measured with a power meter. Dashed lines represent coax connections between the components while solid lines represents waveguide connections. Finally, dash-dot lines represent communication connection (serial or I2C) between components and the RPi.}
    \label{fig:schematic}
\end{figure*}

The core element of PROTOCALC is the calibration source. For this project, we chose a radio-frequency (RF) source that emits only in the frequency band we want to calibrate. As mentioned before, the output frequency of PROTOCALC is in the W-band, so between $\SI{75}{\giga\hertz}$ and $\SI{115}{\giga\hertz}$. For the specific application of PROTOCALC, we set the output frequency at \SI{90}{\giga\hertz}. For the source to produce this signal, we use a Valon-5019 frequency synthesizer that generates a signal up to \SI{20}{\giga\hertz}. This signal is then multiplied by a factor of six by the Eravant frequency multiplier\footnote{\url{https://www.eravant.com/75-to-110-ghz-13-dbm-output-power-wr-10-waveguide-w-band-x6-active-frequency-multiplier}} to the desired frequency. The multiplier waveguide output is in rectangular waveguide (WR10.0) with a spectral range of 75-110 GHz (W-band). As a result, for a typical $30 \%$ bandwidth of CMB telescopes centered around \SI{90}{\giga\hertz}, PROTOCALC can calibrate the whole band. At the output of the multiplier, we have a directional coupler with a \SI{20}{\decibel} directivity and \SI{20}{\decibel} coupling. We decided to use this component so that we can monitor the status of the source with a diode that is read by an Analog-to-Digital Converter (ADC). Even though the presence of the directional coupler attenuates the signal, the multiplier has a fixed output of \SI{13}{\dBm}, so it is necessary to further attenuate the signal. Indeed, also considering the distance of the source from the telescope, this high-power would completely saturate the detectors. So, to reduce the output power, we employ a passive, \SI{30}{\decibel} attenuator after the multiplier. A near-field probe is used as a low-gain (\SI{6.5}{\decibel}i) antenna, which provides a beam of $\simeq$\qtyproduct[product-units=power]{115 x 60}{\SIUnitSymbolDegree} at \SI{90}{\giga\hertz}. The signal emitted by the near-field probe passes through a polarizing wire grid (located at \SI{6}{\milli\meter} after the antenna aperture) that defines the final source module polarization direction and mode-shape (see Ludwig’s first polarization definition \citep{Ludwig1973}).
The grid is manufactured by PureWave Polarizers\footnote{\url{https://purewavepolarizers.com/wire-grid-polarizers/10-micron-wire-far-ir-thz-polarizer}} and uses tungsten wires with \SI{10}{\micro\meter} diameter and wire-spacing of \SI{100}{\micro\meter}. A full representation of the configuration is presented in Figure \ref{fig:schematic}. All the components are controlled by a Raspberry Pi (RPi) unit, specifically the RPi 4b. This computer sets the correct parameters for the frequency, power output, and frequency modulation of the Valon-5019. The modulation phase is estimated using the clock of the RPi. This one is synchronized with GPS time, which is the same timing source for the telescopes.  
An ADC that digitizes the signal output of the diode is connected to the RPi. Finally, this signal is then recorded and time-tagged on the RPi on-board memory for post-processing analysis. 

\subsection{Attitude Determination}\label{subsec:attitude}

\begin{table}
	\centering
	\begin{tabular}{lr}
		Property & Value \\
		\hline
		Field-of-View (FoV) & \SI{84}{\degree} \\
		Maximum Resolution (Video Mode) & \SI{4}{\kelvin} \\
		Maximum frame rate & $\simeq$ \SI{30}{\hertz} \\
		\hline
	\end{tabular}
	\caption{Main properties of the Sony RX0-MII camera.}
	\label{tab:camera}
\end{table}

Achieving a precision of $0.1\si{\SIUnitSymbolDegree}$ requires meticulous attention to the calibration device's orientation and position. As mentioned in Section \ref{UAV_platform}, the DJI drone is equipped with RTK technology, allowing for vertical location accuracy of approximately $\simeq$ \SI{2}{\cm}. While the GPS system provides positional information and rudimentary heading accuracy, the primary goal for PROTOCALC remains to calibrate the instrument's polarization angle. To this end, we have outfitted the payload with our primary sensor for estimating the Euler angles: a photogrammetry camera described in detail in \cite{Dunner2024}. Specifically, we have chosen the Sony RX0-MII. This camera is powered by its internal battery and additionally has its own storage configuration. Indeed, the camera uses an internal microSD to save the videos. The main characteristics of the camera are presented in \ref{tab:camera}. Finally, the camera interfaces with the RPi, as detailed in \ref{app:software}. Furthermore, our approach to attitude reconstruction via photogrammetry involves deploying geo-referenced targets on the ground and analyzing the captured videos to accurately estimate the payload's Euler angles. 

\section{Laboratory Calibration} \label{sec:laboratory}

\begin{table}[]
    \centering
    \begin{tabular}{lcc}
         & Valon Only & Valon + Multiplier \\
         \hline
         Frequency Accuracy & $ \SI{1}{\kilo\hertz}$ & $ \SI{10}{\kilo\hertz}$ \\ 
         Frequency Stability & $ \SI{0.5}{\kilo\hertz}$ & $ \SI{5}{\kilo\hertz}$ \\ 
         Bandwidth &  $ \SI{3}{\kilo\hertz}$ & $ \SI{100}{\kilo\hertz}$ \\
         First Harmonic Level & $\SI{-20}{\decibel}$ & $\SI{-40}{\decibel}$\\
         Output Power & $-3^{*}\si{\dBm}$ & $\SI{12.1}{\dBm}$ \\
         Power Stability & $\SI{-0.01}{\dB}$ & $\SI{0.05}{\dB}$ \\
         \hline
    \end{tabular}
    \caption{Measured properties of the Valon5019 and the Valon and Eravant Multiplier. The properties are measured at nominal output frequency of $\SI{15}{\giga\hertz}$ and \SI{90}{\giga\hertz} for the Valon and Valon + Multiplier, respectively. The output power of the Valon only is set at $\SI{-3}{\dBm}$ due to input power requirements of the multiplier, but can be set at multiple levels. For frequency accuracy, we mean the difference between the measured frequency and the one that has been set.}
    \label{tab:rf_results}
\end{table}

\subsection{RF Calibration} \label{subsec:lab_rfcal}
We first test all RF components separately and then add one component at a time to compose the final RF configuration. The passive components (waveguide directional coupler and fixed attenuator) were analyzed using a Vector Network Analyzer (VNA) to verify specified performance prior to integration. However, for the active components, such as the Valon-5019 and the frequency multiplier, we performed multiple tests using a signal/spectrum analyzer. 

The Valon-5019 performance was assessed in the first testing phase. In particular, we tested the frequency output stability and accuracy of the synthesizer and output power. We tested the Valon for frequencies between \SI{10}{\giga\hertz} and \SI{19}{\giga\hertz}. \footnote{With the new firmware update, the Valon-5019 can achieve a maximum of \SI{20}{\giga\hertz}, however this firmware was not available at the time of testing.} We then tested the Valon coupled to the Multiplier to verify the expected output in the CMB calibration band at \SI{90}{\giga\hertz}. Both tests, Valon only and Valon + Multiplier, were performed using a spectrum analyzer. The model available is limited to \SI{40}{\giga\hertz} input, so for the Valon + Multiplier test we needed to couple the source to the spectrum analyzer using a Harmonic Mixer for the full chain tests.

The test results are presented in Table \ref{tab:rf_results}. Note that these tests were performed using a spectrogram for only five minutes, except for the stability tests that lasted approximately one hour. The analysis indicated that the frequency output and power of the chain neither drift throughout an entire acquisition nor significantly change. Both short and long tests were performed when the multiplier thermalizes, which is between 5 and 10 minutes after turning it on depending on the heat sink and on the environment. We repeated these tests multiple times under different environmental conditions (laboratory temperature varying between 18 and \SI{22}{\celsius}) without noticing any deviation from the values presented here. 

Finally, we also characterize the diode using a power meter and the directional coupler. In particular, we wanted to measure the responsivity of the detector and optimize the position of the directional coupler in the RF chain. Since the output of the multiplier is too large for the telescope, we need to add an attenuator. We position this attenuator before or after the directional coupler. The only limiting factor in this case is the capability of the detector to measure the power at the coupled port of the coupler itself. 
We find that if we position the attenuator before the directional coupler, the diode will receive too little power and will not be able to detect any signal. Consequently, we place the attenuator at the output of the directional coupler to have the diode capable of measuring the signal. In this regime, the responsivity of the diode is measured as \SI{689.5}{\milli\volt / \milli\watt}.

The directional coupler and the attenuator were tested with a VNA and did not show any deviation from the component specification values. 

\begin{figure}
    \centering
    \includegraphics[width=0.9\linewidth, trim={2cm 0 2cm 0}]{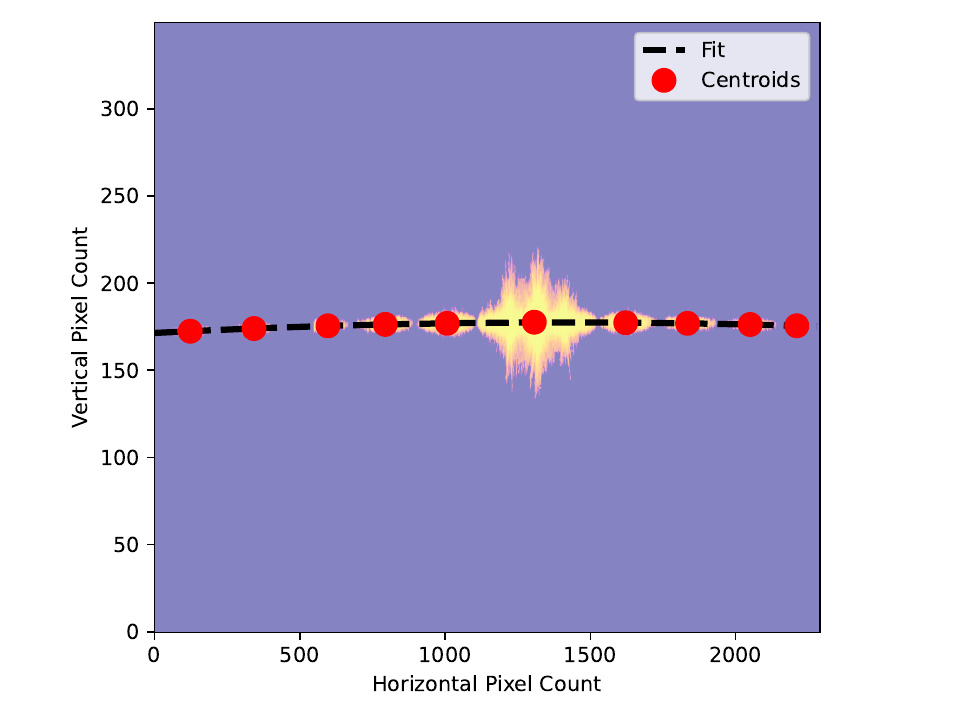}
    \caption{Fitted parabola on the diffraction pattern. The centroids of each diffraction maxima are plotted as red dots.}
    \label{fig:fit}
\end{figure}

\subsection{Optical and Attitude} \label{subsec:lab_opt_att}

The first part of the optical calibration for PROTOCALC involves the intrinsic matrix and distortion coefficients of the camera. For this step, we take advantage of existing software that can be  found in openCV \citep{opencv_library}. We assume a `chessboard-like' calibration pattern at a distance slightly greater than the hyper focal distance of the camera, which for the Sony RX0-II is around $1.7 \si{\meter}$. The calibration pattern size has been chosen so that it fills approximately $60 \%$ of the FoV at a distance of $2 \si{\meter}$. In order to calibrate the camera for operational conditions, we set the camera with the parameters that we plan to use. In particular, we set the recording mode of the camera to manual video mode at resolution of $4096\times2160$ with a framerate of \SI{29.997}{\hertz}. The manual mode ensures that we can set the focus distance to infinity and disable the electronic stabilization. We record a ten-second video and then the code, using openCV routines, first recognizes the pattern in each frame and later uses the corresponding calibration camera function to obtain the camera parameters. The camera parameters describe the camera following the pin-hole camera modeling with modification to include the radial and tangential distortion introduced by the lens \citep{Forsyth2003}. This assumption is valid given the size of the FoV. 

\begin{figure}
	\centering
	\includegraphics[width=0.9\linewidth]{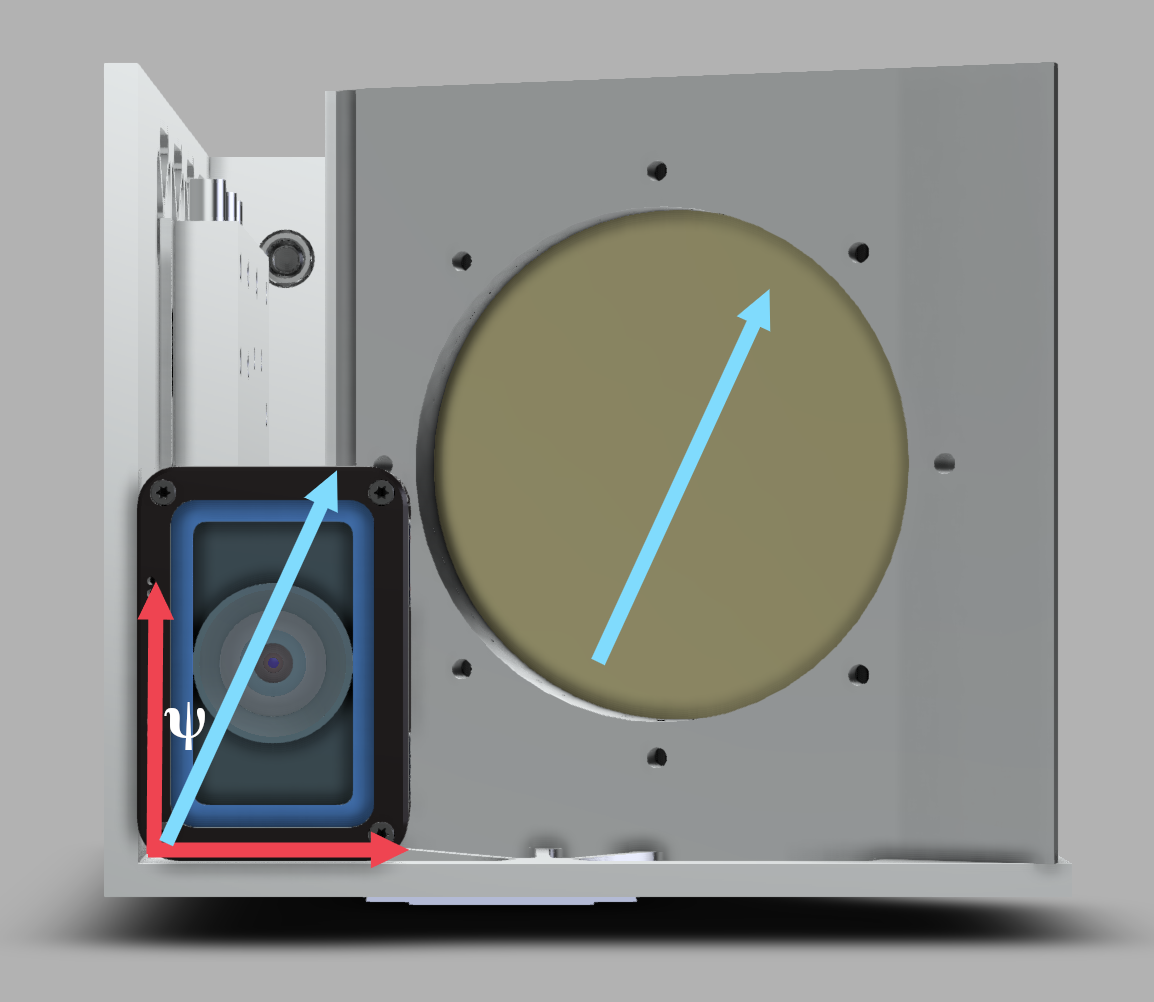}
	\caption{Definition of the rotation angle $\psi$ between the camera and the polarization grid. In PROTOCALC, the grid present a vertical polarization axis, however for visual representation we show a rotated axis to highlight the angle.}
	\label{fig:align}
\end{figure}

Once the camera is calibrated, it is crucial to study its alignment with respect to the polarizing grid as the latter defines the polarization vector of the source, while the camera is used for attitude reconstruction. The relative rotation between the vertical axis of polarizing grid with respect to the vertical axis of the camera is called $\psi$. The geometry is presented in figure \ref{fig:align}.  For this study, we use a laser pointing at the center of the polarizing grid, which creates a diffraction pattern in the far-field. To keep the laser pointed at the center of the grid, we used a 3D-printed mount that is connected to the WR-10 flange of the multiplier. With this method, the calibration can also be performed on the field as long as a flat non-reflective screen is present. 

The diffraction pattern is photographed with the camera, and the images are analyzed. Firstly, the images are pre-processed using a gaussian smoothing and to recognize the diffraction maxima. To obtain the rotation angle, we fit a rotated parabola using two different datasets: one using only the centroid of each blob in the diffraction pattern and one using all the centroids along a column. The choice to fit a parabola is because this curve encompass the information that the grid is tilted with respect to the laser.  Additionally, we used the least-squares and MCMC fitting techniques to compare the results. All the analysis is performed using a laboratory version of \textit{Image Processing Analysis}, IPA \citep{zenodo.15482064}, that has been developed specifically to analyze the images of the project. In all methods, the centroid of the main maximum is used as a prior for the parabola vertex. The resulting fitted parabola is presented in Figure \ref{fig:fit}.

The fitting results are consistent between the methods, and we estimate the error on the angle of a single image as $\Delta\Psi = 0.007\SIUnitSymbolDegree$. When analyzing multiple images, the statistical error reduces to $\Delta\Psi = 0.002\SIUnitSymbolDegree$. The resulting offset between the camera and the grid is $\Psi = 0.03\SIUnitSymbolDegree$. 

The final component optically characterized was the angular response of the near-field probe. The E- and H- planes of the antenna power pattern were measured as a function of angle using the VNA and then the FWHM (Full-Width-Half-Maximum) was derived from the response. We find slightly different values at \SI{90}{\giga\hertz} between fitted and expected FWHM. In particular, we measured \SI{107}{\SIUnitSymbolDegree} and \SI{66}{\SIUnitSymbolDegree} for the for the E- and H- plane beam widths, respectively. This means a difference of a few degrees with respect to the nominal values of \SI{115}{\SIUnitSymbolDegree} and \SI{60}{\SIUnitSymbolDegree}. However, given the large beam on both planes, this is not expected to significantly affect our calibration strategy. 

\section{PROTOCALC Simulation} \label{sec:simulation}

\begin{figure*}
    \centering
    \includegraphics[width=\linewidth]{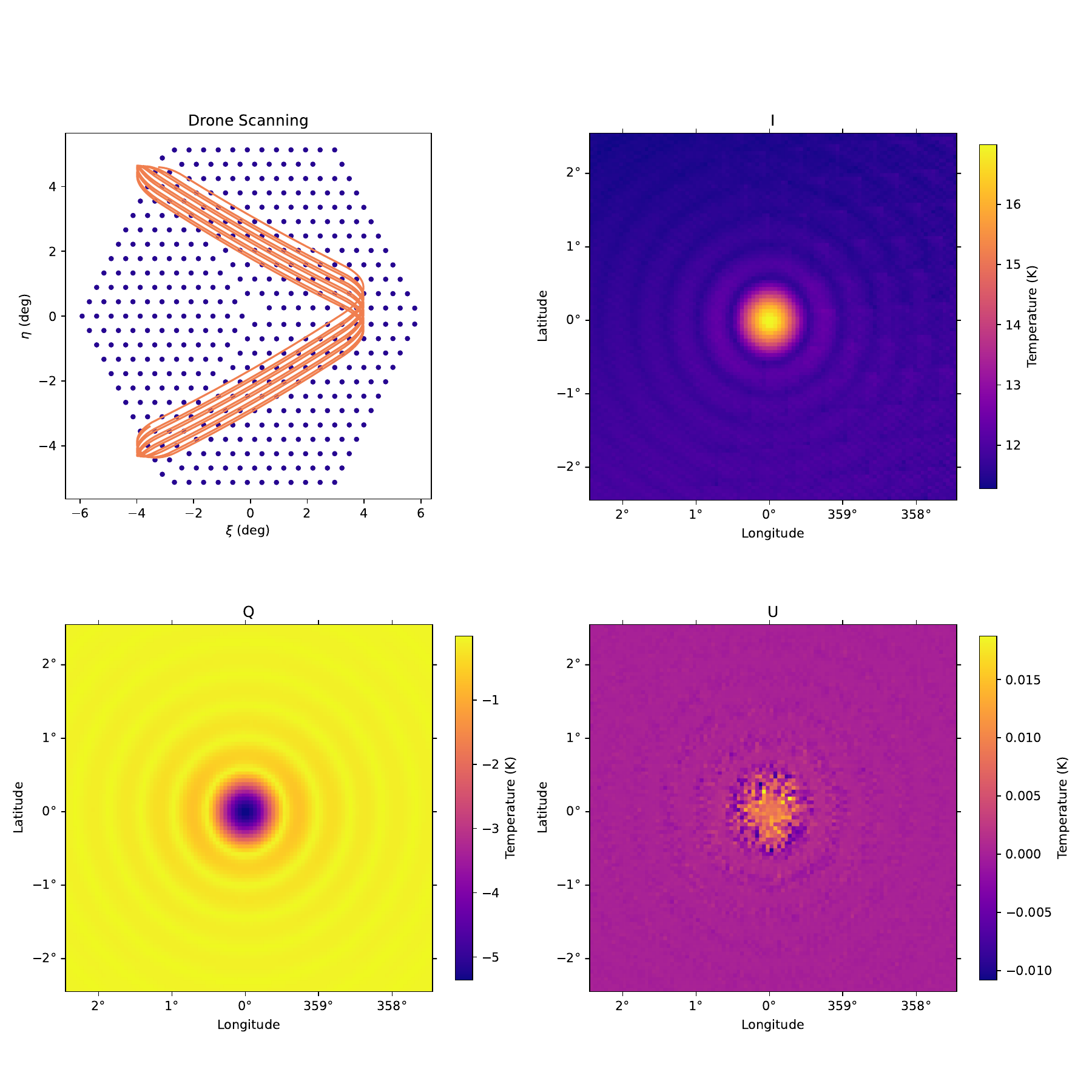}
    \caption{Scanning pattern of the drone over a single array and the resulting maps from this simulated flight. The pattern replicates the programmed flight plan that we have during a typical calibration campaign.}
    \label{fig:simulation}
\end{figure*}

Scheduling the flight and optimizing the calibration campaign were critical tasks for a project of this nature, given that the available calibration time is limited to less than \SI{10}{\minute}. This is due to the UAV’s flight duration at an observing altitude of \SI{5200}{\meter}. To address this challenge, we developed a simulation suite that is based on the \texttt{TOAST} framework \citep{TOAST}. This allows us to generate a time-domain simulation that mimic the calibration campaigns performed in Chile and includes both the environmental and instrumentation effects. The goal was to generate a scan strategy that roughly covers one SO-SAT detector module spanning approximately \SI{12}{\degree} on the sky (the SO-SAT focal plane is comprised of seven wafers in total) and analyze the simulated drone signal, with a particular focus on recovering the absolute polarization angle.
After considering different scanning patterns, such as moving the telescope only in one direction or performing a grid scan, we decided to allow the drone to move only in elevation while the telescope only moves in azimuth. In particular, we chose the drone flight schedule to match one of the schedules designed for the previous observation campaign \cite{Coppi2022} and use the resulting scan strategy for every \SI{10}{\minute} simulation we generated. The projected position of the drone on the focal plane is illustrated in the first panel of Figure \ref{fig:simulation}, from which we find that the achieved drone coverage corresponds roughly to a third of the wafer. Alternatively, $\sim$ 300 detectors will see the drone in a single scan (about 250 of them will see it twice).  

We simulate a calibration campaign for the Simons Observatory Small Aperture Telescopes and focus on the central wafer of one of the Mid-Frequency (MF) telescopes. The drone is positioned \SI{500}{\meter} away from the telescope, transmitting a signal with an amplitude of \SI{-18}{\decibel}$\text{m}$, and a frequency center at \SI{90}{\giga\hertz}. We fix the telescope boresight at a \SI{50}{\degree} elevation at all times and include a random error in the drone position that reflects the realistic uncertainty in the GPS measurements. Moreover, we take into account the effects of acceleration and deceleration and eventual wind gusts that can move the drone from its scheduled path. Furthermore, the simulation pipeline includes the emission from the atmosphere, even though we expect the latter to be significantly sub-dominant to the bright thermal emission of the drone. Indeed, for median weather conditions (PWV $\approx$ \SI{1}{\milli\meter}) at \SI{5200}{\meter} and observing elevation of \SI{50}{\degree} through \SI{500}{\meter} of atmosphere, the drone signal is simulated to be 50 to 60 times brighter than the background atmospheric signal. Therefore, the performance of the simulation analysis is not expected to vary significantly from one wafer to another. We simulate a \SI{10}{minute} flight per day for each day spanning from the beginning of April until the end of July (four months), assuming an azimuth speed of \SI{0.5}{\degree} and sampling at \SI{37}{\hertz}. The SO-SATs Half-Wave-Plate (HWP) is assumed to be continuously spinning during the simulations with a rotation frequency equal to \SI{2}{\hertz}.

As the primary objective of PROTOCALC is to calibrate the polarization angle, we make the simulated signal fully polarized and we assign to each simulation a different value for the polarization angle\footnote{For the purpose of the simulation, the $\Psi$ angle defined in the previous section is identical to the polarization angle. This implies that the passage from the camera coordinate system to the one defined by the grid is already performed.} which is drawn from a Gaussian distribution of mean and standard deviation \SI{90}{\degree} and \SI{0.1}{\degree}, respectively. The latter matches the true uncertainty we expect for the roll angle of the drone. Once the timestreams are generated, they are fed into the \texttt{TOAST} map-maker, which produces the Stokes $I,Q,U$ beam maps that will be used for the polarization angle determination. The second, third and fourth panels of Figure \ref{fig:simulation} show an example beam map set for a single flight. Note that, in practice, the drone data will also be used for pointing calibration and are expected to play a key role in reconstructing the SO-SAT polarized beams. However, for the purposes of this section, we will focus solely on the polarization angle. Results from past drone campaigns on pointing and beam fitting will be presented in an upcoming publication in the near future. Each flight with this scanning strategy covers $\simeq 100$ detectors with at least 3 crossing per detector, the SNR per detector is expected above \SI{40}{\decibel} \cite{Dachlythra_2024}.

Once the $\sim 120$ simulations are mapped, we estimate the absolute polarization angle, $\phi$, by fitting 2D Gaussian beams to each simulation's $Q$ and $U$ beam maps and employing the corresponding fitted amplitudes:
\begin{equation}
\phi = \frac{1}{2}\arctan\left(\frac{U}{Q}\right).
\end{equation}
The achieved reconstruction uncertainty on $\phi$, is estimated as $\sigma \approx 0.098\si{\degree}$. The small reconstruction error further highlights the advantages of using an artificial source with increased polarization brightness, such as PROTOCALC. 

\section{In-Flight Performance} \label{sec:inflight}

\begin{figure*}
    \centering
    \includegraphics[width=\textwidth]{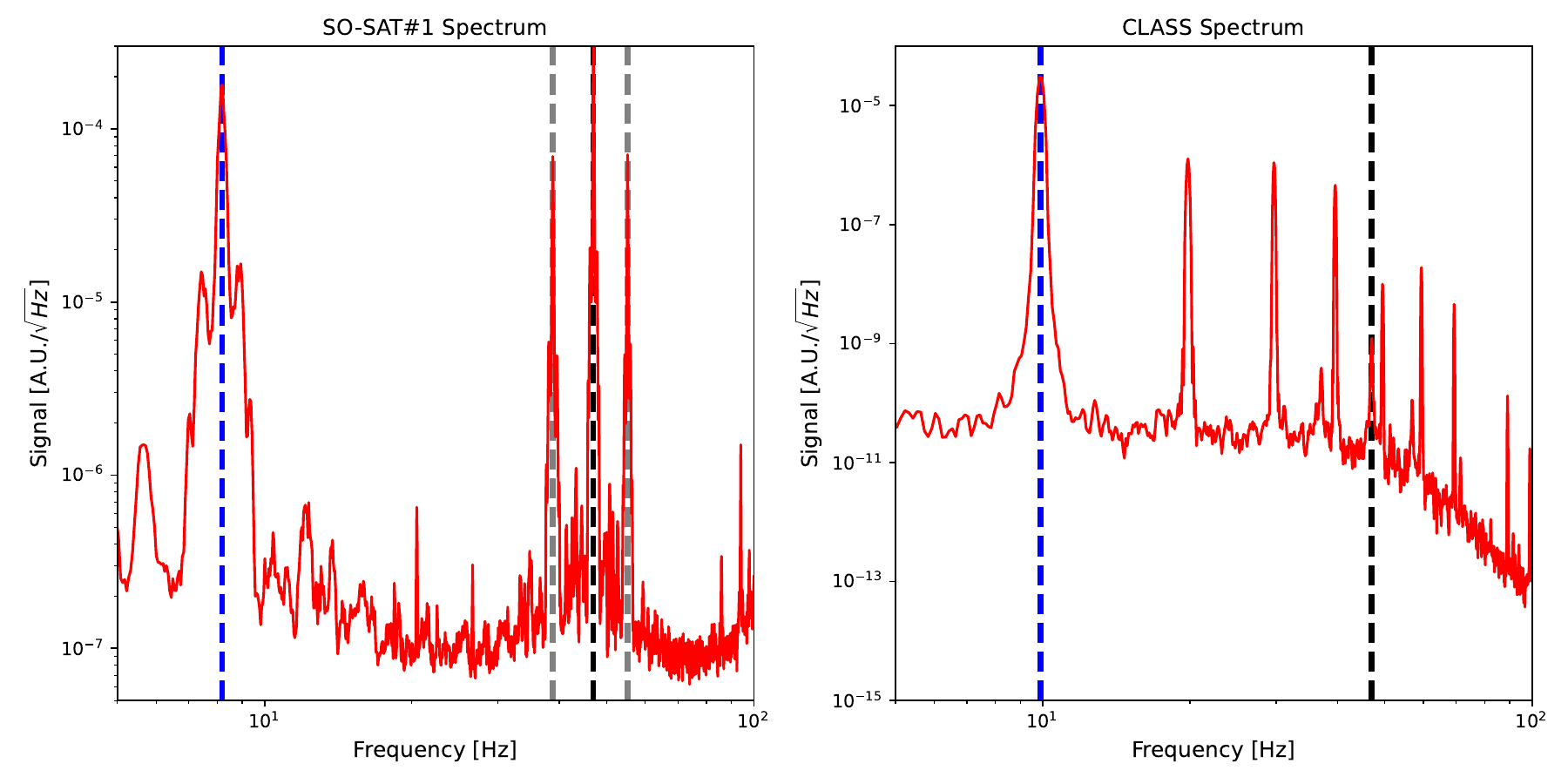}
    \caption{Power spectra of a single detector throughout an entire flight for both Simons Observatory and CLASS. For the SO power spectrum, we highlight both the \SI{8}{\hertz} modulation frequency of the HWP in blue and the $47-$\SI{8}{\hertz} and $47+$\SI{8}{\hertz} in gray. Instead, for CLASS the VPM signal is in blue, notice the small lines at $47-$\SI{10}{\hertz} and $47+$\SI{10}{\hertz} that are present here. For both power spectra, we also highlight in black the \SI{47}{\hertz} signal coming from the source.}
    \label{fig:signal}
\end{figure*}

PROTOCALC has been tested in the field three times. The first time was in April 2022, the second one in January 2023 while in May of 2024. For all these campaigns, PROTOCALC flew along with another source at $150\si{\giga\hertz}$, namely Hovercal \citep{Dunner2024}. 

PROTOCALC was tested while calibrating one of the CLASS telescope (W-band) for all the flights, while Simons Observatory was targeted only for the latest campaign flight. For SO, we targeted 2 of the MF Small Aperture Telescopes looking especially at the platform 1 and 3, from now on SO-SAT1 and SO-SAT3. Throughout a single calibration day, we alternated the two sources to calibrate different bands. We equally split the number of flights between the two payloads, with a maximum number of 6 flights per payload.  Each flight was between $10\si{\minute}$ and $12\si{\minute}$ with approximately $5\si{\minute}$ total between ascending and descending. The drone flew at an altitude of $350\si{\meter}$ with respect to the starting point scanning in elevation and keeping a constant distance of approximately $500\si{\meter}$ from the telescope. With the simulation framework introduced in section \ref{sec:simulation}, we found that the optimal scan strategy for the SO-SATs involves moving the telescope only in Azimuth. 
The RF source was tested in different configurations with and without a chopping on. The frequency chosen for the chopping system was set to $47\si{\hertz}$, which is far from the $10\si{\hertz}$ modulation and away from the higher harmonics of the modulation frequency introduced by the Variable-delay Polarization Modulator, VPM \citep{harrington2018}, system on CLASS and further from the $8\si{\hertz}$ modulation of the SO HWP \citep{yamada2024simons}. The modulation is introduced at the valon level that supports amplitude modulation via firmware. 

A single flight of the drone can target the calibration of multiple telescopes. Specifically, during the third flight both SO-SAT1 and SO-SAT3 observed the drone and the CLASS W-band telescope. For this flight, the source was moving in elevation while pointing at SO-SAT3\footnote{The platform for the SO-SAT3 is located in the middle between CLASS and SO-SAT1 and was chosen for this reason.}, while all three telescopes were moving in azimuth. 

The raw signal, as observed during the same flight from SO-SAT1 and from CLASS W-band, is shown in Figure \ref{fig:signal}. The source signal at the modulation frequency of  $47\si{\hertz}$ is evident in both illustrated timestreams. Further analysis must be performed to extract the signal and produce the calibration maps. However, the presence of the strong line at the chop frequency indicates that the source polarized emission is seen by the telescopes.

\begin{figure}
    \centering
    \includegraphics[width=\linewidth]{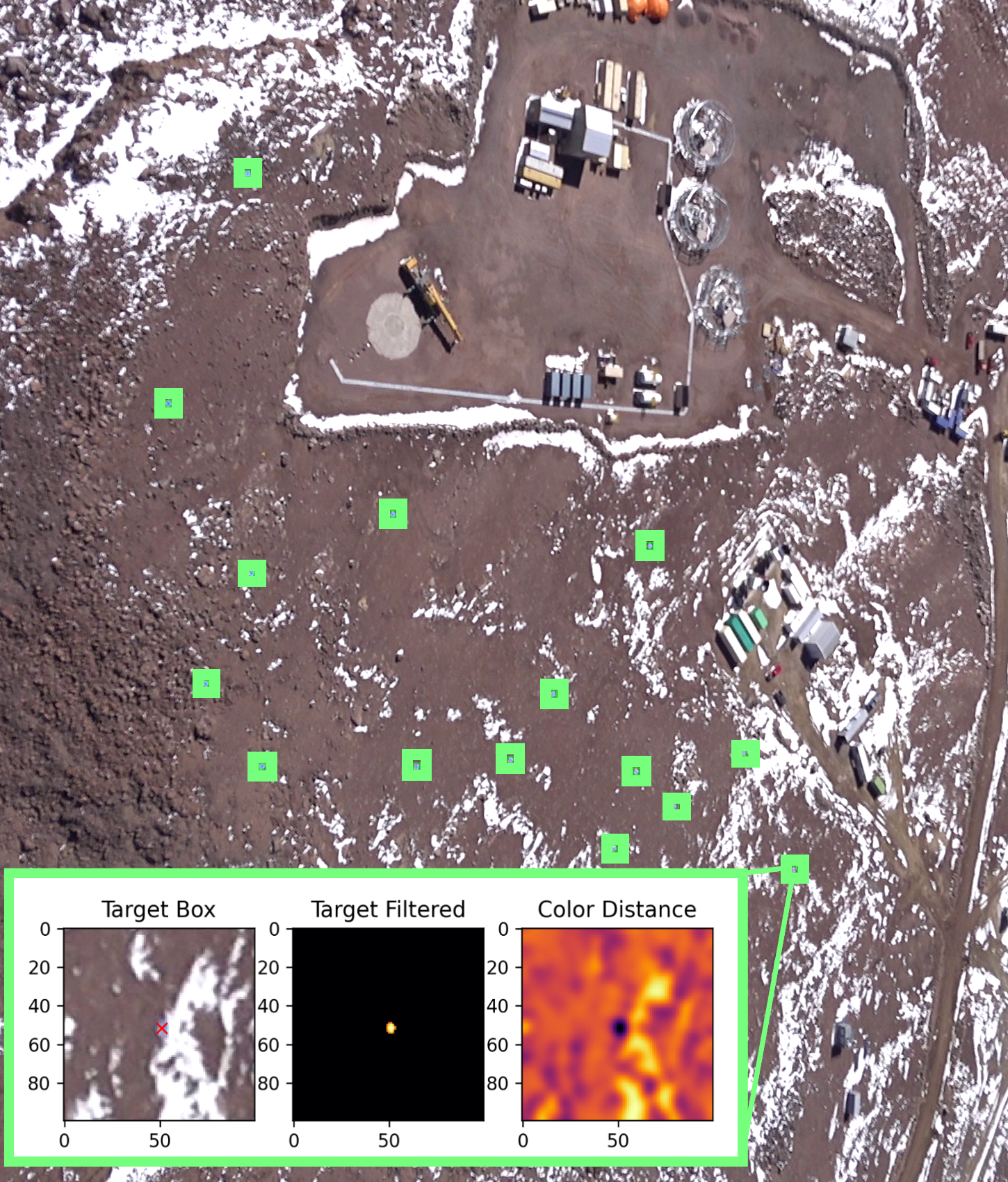}
    \caption{View of the site as seen by the Photogrammetry camera. The recognized targets are enclosed in light green square and a zoom of one of them is presented. Here, we show the process to recognize the target from the image.}
    \label{fig:view}
\end{figure}

As mentioned in Section \ref{subsec:attitude}, to compute the attitude of the source, the main sensor is the photogrammetry camera. The camera is in video mode with the manual focus set to infinity. Other parameters, such as the ISO and shutter speed, are chosen to maximize the visibility of the ground targets. To analyze the video, we developed the flight version of IPA. In this code, the initial step is to locate the target in each video frame. To do this, we start by inputting each target's coordinates in pixel space along with its color in RGB space. To estimate the target location in the second frame, we compute the homography matrix \citep{Hartley2004} between the frames to estimate a preliminary extrinsic matrix and a preliminary new target location. Then, we convert the frame in CIELAB color space, \citep{international2004colorimetry} and we search the target around the preliminary location using as a metric the $\Delta E_{2000}$ color distance \citep{Sharma2005}. The target location is then estimated as the centroid of the filtered distance map. This process is then repeated for the third frame, knowing the position from the second frame and so on until the end of the video. The target recognition and the layout of the targets distribution is presented in Figure \ref{fig:view}. 

\begin{figure}
    \centering
    \includegraphics[width=\linewidth]{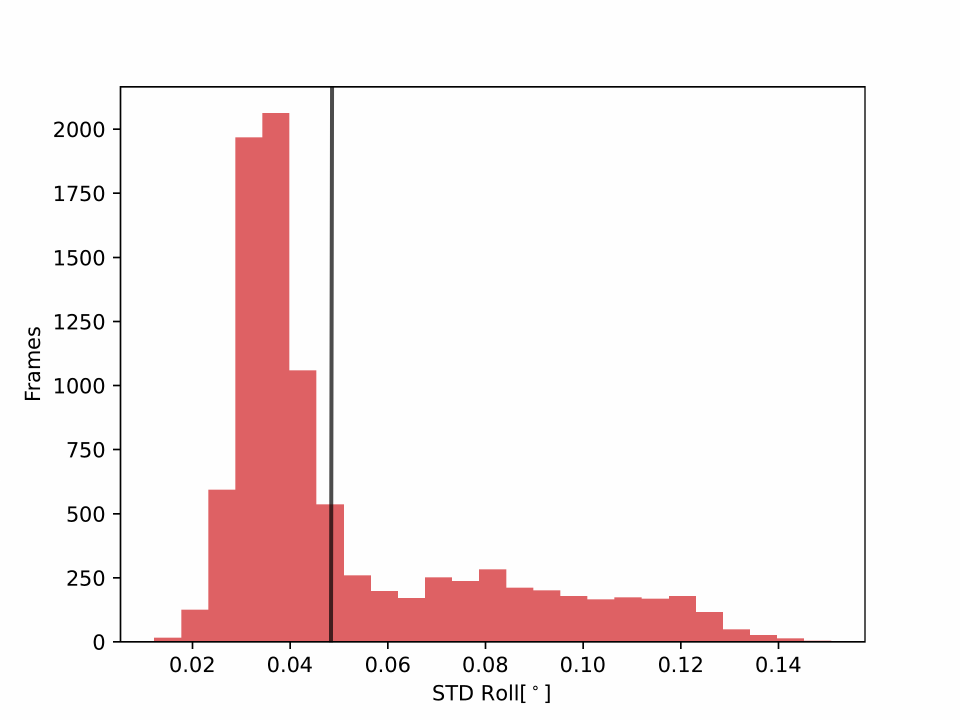}
    \caption{Histogram presenting the uncertainty of on the roll angle of the camera estimated using the Jackknife resampling for each frame. The black vertical line represents the average value of the uncertainty.}
    \label{fig:roll}
\end{figure}

Once we obtain all coordinates in pixel space, we can then estimate the translation and rotation vectors using Perspective-n-Point techniques implemented in openCV. These algorithms are capable of estimating the position of the camera with respect to fixed points in space given a camera model and parameters. Indeed, all the target positions are known with high precision since they have been measured using RTK. However, there are degeneracies between the elements in the translation and rotation vectors. To break this degeneracy, we use the translation vector provided by the GPS measurements and minimize the reprojection error using least-squares. The uncertainty in the reconstructed camera roll angle is presented in Figure \ref{fig:roll}. The roll angle is a proxy for the polarization angle, indeed we need to include the projection effects due to yaw, pitch and the grid rotation to get the latter. These values have been computed using jackknife resampling, where the angles are reconstructed each time, removing one of the targets at the solvePnP level. The mean value is $0.045\si{\SIUnitSymbolDegree}$, which is lower than the designed value accuracy for PROTOCALC. However, it is important to notice some outliers, which correspond mostly to times when the drone moves fast, especially at turn-around points. The results from the roll angle analysis provide crucial information for next calibration campaigns. Indeed, combining this information with simulation, we can choose the best calibration strategy that factors in drone speeds. 

\section{Conclusion}

In this paper we presented a new calibrator for CMB telescopes that has been successfully developed and deployed in the field. The full calibration analysis for the CLASS and SO telescopes is ongoing and it will be presented in full in a future publication. In the laboratory, we show clearly that the source can be aligned with an accuracy smaller than $0.01\SIUnitSymbolDegree$. This shows the potential to use this source for next-generation ground based experiments, such as CMB-S4. We successfully performed calibration tests multiple times at the site in Chile with multiple telescopes, CLASS W-band and two SO-SATs. Throughout these tests, the source was observed by the telescopes. Additionally, we analyze the data from the campaign to show the accuracy of the roll angle reconstruction. In particular we found a mean accuracy value of $0.045\si{\SIUnitSymbolDegree}$ throughout a single flight. This accuracy is currently the limiting factor for the accuracy of PROTOCALC, and future versions of the instrument will further improve this value through the use of additional motion sensors in combination with the photogrammetry. The purpose of the paper was to highlight the potential of PROTOCALC and thus presents only the roll angle. However, we notice that future publications will present the polarization angle calibration for both telescopes using this method and including the effects of the polarization modulators. \newline \newline

\textbf{Acknowledgments}: Co-funded by the European Union (ERC, POLOCALC PI Nati, 101096035 and MSCA, PROTOCALC PI Coppi). Views and opinions expressed are however those of the authors only and do not necessarily reflect those of the European Union or the European Research Council. Neither the European Union nor the granting authority can be held responsible for them. This work was supported in part by a grant from the Simons Foundation (Award $\#$457687, B.K.). \\
This document was prepared by Simons Observatory using the resources of the Fermi National Accelerator Laboratory (Fermilab), a U.S. Department of Energy, Office of Science, Office of High Energy Physics HEP User Facility. Fermilab is managed by Fermi Forward Discovery Group, LLC, acting under Contract No. 89243024CSC000002. JE, Ema Tsang King Sang, Amalia Villarrubia Aguilar acknowledge funding from the SCIPOL project\footnote{\url{scipol.in2p3.fr}} funded by the European Research Council (ERC) under the European Union’s Horizon 2020 research and innovation program (PI: Josquin Errard, Grant agreement No. 101044073). Alexandre Adler is supported by the NSF under grant MSRI-2 2153201)

\appendix

\section{Software Design}\label{app:software}

PROTOCALC utilizes the Raspberry Pi 4 as the main flight computer. This model offers the advantage of multiple serial outputs and a CPU with multiple threads. The nominal power consumption of the RPi4 is approximately \SI{3.5}{\watt}, well within the power budget. PROTOCALC is managed by an in-house developed Python code named PORTER, which stands for \textit{Protocalc cOntRol sysTEm pRogram} \citep{zenodo.15482064}. The software initiates automatically upon powering up the RPi. Initially, it checks the time and GPS signal, updating the RPi's clock using the chrony software\footnote{\url{https://chrony-project.org/index.html}}. In the absence of any GPS signal, an external Real-Time Clock (RTC), maintains accurate time with a precision of $2\text{ppm}$ at operational temperatures\footnote{After every calibration run at the site, the RPi is also connected to internet again to sync the internal clock also against time servers.}. Following clock synchronization, PORTER configures the necessary parameters for the Valon-5019. These two are the highest priority operations and are executed serially as they only need to occur once when beginning the operations. 
After initialization of these core components, PORTER controls and reads multiple sensors, including the Camera, the ADC and the GPS. Due to this multi-sensor configuration, PORTER is threaded to deal with multiple sensors concurrently. Each sensor but the camera uses the same code structure, so that it is possible to create a yaml file with a series of parameters to configure the sensors. The camera requires specific software since it uses the Picture Transfer Protocol, PTP. For this reason we control the camera using the core commands of \textit{SOny gUi Remote}, SOUR \citep{zenodo.15482064}, a Python code designed to manage multiple Sony cameras through a graphical user interface (GUI)\footnote{The GUI is utilized solely in the laboratory for camera calibration purposes.}. Leveraging the RPi's clock, it is feasible to synchronize the camera's timing and adjust its internal clock before commencing video recording. 

\clearpage

\bibliography{main.bib}{}

\begin{thebibliography}{}
\expandafter\ifx\csname natexlab\endcsname\relax\def\natexlab#1{#1}\fi
\providecommand{\url}[1]{\href{#1}{#1}}
\providecommand{\dodoi}[1]{doi:~\href{http://doi.org/#1}{\nolinkurl{#1}}}
\providecommand{\doeprint}[1]{\href{http://ascl.net/#1}{\nolinkurl{http://ascl.net/#1}}}
\providecommand{\doarXiv}[1]{\href{https://arxiv.org/abs/#1}{\nolinkurl{https://arxiv.org/abs/#1}}}

\bibitem[{{Abazajian} {et~al.}(2022){Abazajian}, {Addison}, {Adshead}, {Ahmed}, {Akerib}, {Ali}, {Allen}, {Alonso}, {Alvarez}, {Amin}, {Anderson}, {Arnold}, {Ashton}, {Baccigalupi}, {Bard}, {Barkats}, {Barron}, {Barry}, {Bartlett}, {Basu Thakur}, {Battaglia}, {Bean}, {Bebek}, {Bender}, {Benson}, {Bianchini}, {Bischoff}, {Bleem}, {Bock}, {Bocquet}, {Boddy}, {Richard Bond}, {Borrill}, {Bouchet}, {Brinckmann}, {Brown}, {Bryan}, {Buza}, {Byrum}, {Hervias Caimapo}, {Calabrese}, {Calafut}, {Caldwell}, {Carlstrom}, {Carron}, {Cecil}, {Challinor}, {Chang}, {Chinone}, {Sherry Cho}, {Cooray}, {Coulton}, {Crawford}, {Crites}, {Cukierman}, {Cyr-Racine}, {de Haan}, {Delabrouille}, {Devlin}, {Di Valentino}, {Dierickx}, {Dobbs}, {Duff}, {Dvorkin}, {Eimer}, {Elleflot}, {Errard}, {Essinger-Hileman}, {Fabbian}, {Feng}, {Ferraro}, {Filippini}, {Flauger}, {Flaugher}, {Fraisse}, {Frolov}, {Galitzki}, {Gallardo}, {Galli}, {Ganga}, {Gerbino}, {Gluscevic}, {Goeckner-Wald}, {Green}, {Grin}, {Grohs}, {Gualtieri}, {Gudmundsson},
  {Gullett}, {Gupta}, {Habib}, {Halpern}, {Halverson}, {Hanany}, {Harrington}, {Hasegawa}, {Hasselfield}, {Hazumi}, {Heitmann}, {Henderson}, {Hensley}, {Hill}, {Colin Hill}, {Hlo{\v{z}}ek}, {Patty Ho}, {Hoang}, {Holder}, {Holzapfel}, {Hood}, {Hubmayr}, {Huffenberger}, {Hui}, {Irwin}, {Jeong}, {Johnson}, {Jones}, {Hwan Kang}, {Karkare}, {Katayama}, {Keskitalo}, {Kisner}, {Knox}, {Koopman}, {Kosowsky}, {Kovac}, {Kovetz}, {Kuhlmann}, {Kuo}, {Kusaka}, {L{\"a}hteenm{\"a}ki}, {Lawrence}, {Lee}, {Lewis}, {Li}, {Linder}, {Loverde}, {Lowitz}, {Lubin}, {Madhavacheril}, {Mantz}, {Marques}, {Matsuda}, {Mauskopf}, {McCarrick}, {McMahon}, {Daniel Meerburg}, {Melin}, {Menanteau}, {Meyers}, {Millea}, {Mohr}, {Moncelsi}, {Monzani}, {Mroczkowski}, {Mukherjee}, {Nagy}, {Namikawa}, {Nati}, {Natoli}, {Newburgh}, {Niemack}, {Nishino}, {Nord}, {Novosad}, {O'Brient}, {Padin}, {Palladino}, {Partridge}, {Petravick}, {Pierpaoli}, {Pogosian}, {Prabhu}, {Pryke}, {Puglisi}, {Racine}, {Rahlin}, {Sathyanarayana Rao}, {Raveri}, {Reichardt},
  {Remazeilles}, {Rocha}, {Roe}, {Roy}, {Ruhl}, {Salatino}, {Saliwanchik}, {Schaan}, {Schillaci}, {Schmitt}, {Schmittfull}, {Scott}, {Sehgal}, {Shandera}, {Sherwin}, {Shirokoff}, {Simon}, {Slosar}, {Spergel}, {St. Germaine}, \& {Staggs}}]{s4collaboration2020cmbs4}
{Abazajian}, K., {Addison}, G.~E., {Adshead}, P., {et~al.} 2022, \apj, 926, 54, \dodoi{10.3847/1538-4357/ac1596}

\bibitem[{Abitbol {et~al.}(2021)Abitbol, Alonso, Simon, Lashner, Crowley, Ali, Azzoni, Baccigalupi, Barron, Brown, \& et~al.}]{Abitbol_2021}
Abitbol, M.~H., Alonso, D., Simon, S.~M., {et~al.} 2021, Journal of Cosmology and Astroparticle Physics, 2021, 032, \dodoi{10.1088/1475-7516/2021/05/032}

\bibitem[{{Ade} {et~al.}(2025){Ade}, {Ahmed}, {Amiri}, {Barkats}, {Basu Thakur}, {Bischoff}, {Beck}, {Bock}, {Boenish}, {Buza}, {Cheshire}, {Connors}, {Cornelison}, {Crumrine}, {Cukierman}, {Denison}, {Duband}, {Eiben}, {Elwood}, {Fatigoni}, {Filippini}, {Fortes}, {Gao}, {Giannakopoulos}, {Goeckner-Wald}, {Goldfinger}, {Grayson}, {Greathouse}, {Grimes}, {Hall}, {Halal}, {Halpern}, {Hand}, {Harrison}, {Henderson}, {Hubmayr}, {Hui}, {Irwin}, {Kang}, {Karkare}, {Kefeli}, {Kovac}, {Kuo}, {Lau}, {Lautzenhiser}, {Lennox}, {Liu}, {Megerian}, {Minutolo}, {Moncelsi}, {Nakato}, {Nguyen}, {O'Brient}, {Patel}, {Petroff}, {Polish}, {Prouve}, {Pryke}, {Reintsema}, {Romand}, {Salatino}, {Schillaci}, {Schmitt}, {Singari}, {Sjoberg}, {Soliman}, {St Germaine}, {Steiger}, {Steinbach}, {Sudiwala}, {Thompson}, {Tsai}, {Tucker}, {Turner}, {Verg{\`e}s}, {Vieregg}, {Wandui}, {Weber}, {Willmert}, {Wu}, {Yang}, {Yu}, {Zeng}, {Zhang}, {Zhang}, \& {(Bicep/}}]{Bicep2025}
{Ade}, P.~A.~R., {Ahmed}, Z., {Amiri}, M., {et~al.} 2025, \prd, 111, 063505, \dodoi{10.1103/PhysRevD.111.063505}

\bibitem[{{Aumont, J.} {et~al.}(2020){Aumont, J.}, {Macías-Pérez, J. F.}, {Ritacco, A.}, {Ponthieu, N.}, \& {Mangilli, A.}}]{Aumont_2021}
{Aumont, J.}, {Macías-Pérez, J. F.}, {Ritacco, A.}, {Ponthieu, N.}, \& {Mangilli, A.} 2020, \aap, 634, A100, \dodoi{10.1051/0004-6361/201833504}

\bibitem[{{{BICEP}2/Keck Array {XI}} {et~al.}(2019){{BICEP}2/Keck Array {XI}}, Ade, Ahmed, Aikin, Barkats, Benton, Bischoff, Bock, Bowens-Rubin, Brevik, Buder, Bullock, Buza, Connors, Cornelison, Crill, Crumrine, Dierickx, Duband, Filippini, Fliescher, Grayson, Hall, Halpern, Harrison, Hildebrandt, Hilton, Hui, Irwin, Kang, Karkare, Karpel, Kaufman, Keating, Kefeli, Kernasovskiy, Kovac, Kuo, Larsen, Lau, Leitch, Lueker, Megerian, Moncelsi, Namikawa, Netterfield, Nguyen, O'Brient, IV, Palladino, Pryke, Racine, Richter, Schillaci, Schwarz, Sheehy, Soliman, Germaine, Staniszewski, Steinbach, Sudiwala, Teply, Thompson, Tolan, Tucker, Turner, Umilta, Vieregg, Wandui, Weber, Wiebe, Willmert, Wong, Wu, Yang, Yoon, \& and}]{BK2019}
{{BICEP}2/Keck Array {XI}}, Ade, P. A.~R., Ahmed, Z., {et~al.} 2019, The Astrophysical Journal, 884, 114, \dodoi{10.3847/1538-4357/ab391d}

\bibitem[{{BICEP/Keck Collaboration XIII} {et~al.}(2021){BICEP/Keck Collaboration XIII}, Ade, Ahmed, Amiri, Barkats, Thakur, Bischoff, Beck, Bock, Boenish, Bullock, \& et~al.}]{bicep_keck_r_constraints_2021}
{BICEP/Keck Collaboration XIII}, Ade, P., Ahmed, Z., {et~al.} 2021, Physical Review Letters, 127, \dodoi{10.1103/physrevlett.127.151301}

\bibitem[{Bradski(2000)}]{opencv_library}
Bradski, G. 2000, Dr. Dobb's Journal of Software Tools

\bibitem[{Carroll(1998)}]{Carrol1998}
Carroll, S.~M. 1998, Phys. Rev. Lett., 81, 3067, \dodoi{10.1103/PhysRevLett.81.3067}

\bibitem[{Carroll {et~al.}(1990)Carroll, Field, \& Jackiw}]{Carrol1990}
Carroll, S.~M., Field, G.~B., \& Jackiw, R. 1990, Phys. Rev. D, 41, 1231, \dodoi{10.1103/PhysRevD.41.1231}

\bibitem[{{Coppi}(2025)}]{zenodo.15482064}
{Coppi}, G. 2025, PROTOCALC Code,  Zenodo, \dodoi{10.5281/zenodo.15482064}

\bibitem[{{Coppi} {et~al.}(2022){Coppi}, {Conenna}, {Savorgnano}, {Carrero}, {D{\"u}nner Planella}, {Galitzki}, {Nati}, \& {Zannoni}}]{Coppi2022}
{Coppi}, G., {Conenna}, G., {Savorgnano}, S., {et~al.} 2022, in Society of Photo-Optical Instrumentation Engineers (SPIE) Conference Series, Vol. 12190, Millimeter, Submillimeter, and Far-Infrared Detectors and Instrumentation for Astronomy XI, ed. J.~{Zmuidzinas} \& J.-R. {Gao}, 1219015, \dodoi{10.1117/12.2628312}

\bibitem[{Dachlythra {et~al.}(2024)Dachlythra, Duivenvoorden, Gudmundsson, Hasselfield, Coppi, Adler, Alonso, Azzoni, Chesmore, Fabbian, Ganga, Gerras, Jaffe, Johnson, Keating, Keskitalo, Kisner, Krachmalnicoff, Lungu, Matsuda, Naess, Page, Puddu, Puglisi, Simon, Teply, Tsan, Wollack, Wolz, \& Xu}]{Dachlythra_2024}
Dachlythra, N., Duivenvoorden, A.~J., Gudmundsson, J.~E., {et~al.} 2024, The Astrophysical Journal, 961, 138, \dodoi{10.3847/1538-4357/ad0969}

\bibitem[{{Dahal} {et~al.}(2022){Dahal}, {Appel}, {Datta}, {Brewer}, {Ali}, {Bennett}, {Bustos}, {Chan}, {Chuss}, {Cleary}, {Couto}, {Denis}, {D{\"u}nner}, {Eimer}, {Espinoza}, {Essinger-Hileman}, {Golec}, {Harrington}, {Helson}, {Iuliano}, {Karakla}, {Li}, {Marriage}, {McMahon}, {Miller}, {Novack}, {N{\'u}{\~n}ez}, {Osumi}, {Padilla}, {Palma}, {Parker}, {Petroff}, {Reeves}, {Rhoades}, {Rostem}, {Valle}, {Watts}, {Weiland}, {Wollack}, \& {Xu}}]{Dahal2022}
{Dahal}, S., {Appel}, J.~W., {Datta}, R., {et~al.} 2022, \apj, 926, 33, \dodoi{10.3847/1538-4357/ac397c}

\bibitem[{{de la Hoz} {et~al.}(2022){de la Hoz}, {Diego-Palazuelos}, {Mart{\'\i}nez-Gonz{\'a}lez}, {Vielva}, {Barreiro}, \& {Bilbao-Ahedo}}]{delahoz2022}
{de la Hoz}, E., {Diego-Palazuelos}, P., {Mart{\'\i}nez-Gonz{\'a}lez}, E., {et~al.} 2022, \jcap, 2022, 032, \dodoi{10.1088/1475-7516/2022/03/032}

\bibitem[{Duivenvoorden {et~al.}(2019)Duivenvoorden, Gudmundsson, \& Rahlin}]{beamconv_2018}
Duivenvoorden, A.~J., Gudmundsson, J.~E., \& Rahlin, A.~S. 2019, Monthly Notices of the Royal Astronomical Society, 486, 5448–5467, \dodoi{10.1093/mnras/stz1143}

\bibitem[{D{\"u}nner {et~al.}(2020)D{\"u}nner, Flux{\'a}, Best, \& Carrero}]{Dunner2020}
D{\"u}nner, R., Flux{\'a}, J., Best, S., \& Carrero, F. 2020, in Millimeter, Submillimeter, and Far-Infrared Detectors and Instrumentation for Astronomy X, ed. J.~Zmuidzinas \& J.-R. Gao, Vol. 11453, International Society for Optics and Photonics (SPIE), 114532P, \dodoi{10.1117/12.2561165}

\bibitem[{D{\"u}nner {et~al.}(2024)D{\"u}nner, Astori, Coppi, Eimer, Errard, Flux{\'a}, Li, Mezzanzanica, Nati, Petroff, Puddu, Rojas, \& Zannoni}]{Dunner2024}
D{\"u}nner, R., Astori, F., Coppi, G., {et~al.} 2024, in Millimeter, Submillimeter, and Far-Infrared Detectors and Instrumentation for Astronomy XII, ed. J.~Zmuidzinas \& J.-R. Gao, Vol. 13102, International Society for Optics and Photonics (SPIE), 1310217, \dodoi{10.1117/12.3018469}

\bibitem[{Dutcher {et~al.}(2021)Dutcher, Balkenhol, Ade, Ahmed, Anderes, Anderson, Archipley, Avva, Aylor, Barry, Basu~Thakur, Benabed, Bender, Benson, Bianchini, Bleem, Bouchet, Bryant, Byrum, Carlstrom, Carter, Cecil, Chang, Chaubal, Chen, Cho, Chou, Cliche, Crawford, Cukierman, Daley, de~Haan, Denison, Dibert, Ding, Dobbs, Everett, Feng, Ferguson, Foster, Fu, Galli, Gambrel, Gardner, Goeckner-Wald, Gualtieri, Guns, Gupta, Guyser, Halverson, Harke-Hosemann, Harrington, Henning, Hilton, Hivon, Holder, Holzapfel, Hood, Howe, Huang, Irwin, Jeong, Jonas, Jones, Khaire, Knox, Kofman, Korman, Kubik, Kuhlmann, Kuo, Lee, Leitch, Lowitz, Lu, Meyer, Michalik, Millea, Montgomery, Nadolski, Natoli, Nguyen, Noble, Novosad, Omori, Padin, Pan, Paschos, Pearson, Posada, Prabhu, Quan, Raghunathan, Rahlin, Reichardt, Riebel, Riedel, Rouble, Ruhl, Sayre, Schiappucci, Shirokoff, Smecher, Sobrin, Stark, Stephen, Story, Suzuki, Thompson, Thorne, Tucker, Umilta, Vale, Vanderlinde, Vieira, Wang, Whitehorn, Wu, Yefremenko, Yoon, \&
  Young}]{Dutcher_2021}
Dutcher, D., Balkenhol, L., Ade, P., {et~al.} 2021, Physical Review D, 104, \dodoi{10.1103/physrevd.104.022003}

\bibitem[{Forsyth \& Ponce(2012)}]{Forsyth2003}
Forsyth, D.~A., \& Ponce, J. 2012, Computer Vision - A Modern Approach, Second Edition. (Pitman), 1--791

\bibitem[{{Galitzki} {et~al.}(2024){Galitzki}, {Tsan}, {Spisak}, {Randall}, {Silva-Feaver}, {Seibert}, {Lashner}, {Adachi}, {Adkins}, {Alford}, {Arnold}, {Ashton}, {Austermann}, {Baccigalupi}, {Bazarko}, {Beall}, {Bhimani}, {Bixler}, {Coppi}, {Corbett}, {Crowley}, {Crowley}, {Day-Weiss}, {Devlin}, {Dicker}, {DiGia}, {Dow}, {Duell}, {Duff}, {Gerras}, {Groh}, {Gudmundsson}, {Harrington}, {Hasegawa}, {Healy}, {Henderson}, {Hubmayr}, {Iuliano}, {Johnson}, {Keating}, {Keller}, {Kiuchi}, {Kofman}, {Koopman}, {Kusaka}, {Lee}, {Lew}, {Lin}, {Link}, {Lucas}, {Lungu}, {Mangu}, {McMahon}, {Miller}, {Moore}, {Morshed}, {Nakata}, {Nati}, {Newburgh}, {Nguyen}, {Niemack}, {Page}, {Sakaguri}, {Sakurai}, {Sathyanarayana Rao}, {Saunders}, {Shroyer}, {Sugiyama}, {Tajima}, {Takeuchi}, {Bua}, {Teply}, {Terasaki}, {Ullom}, {Van Lanen}, {Vavagiakis}, {Vissers}, {Walters}, {Wang}, {Xu}, {Yamada}, \& {Zheng}}]{Galitzki2024}
{Galitzki}, N., {Tsan}, T., {Spisak}, J., {et~al.} 2024, \apjs, 274, 33, \dodoi{10.3847/1538-4365/ad64c9}

\bibitem[{Harrington {et~al.}(2018)Harrington, Eimer, Chuss, Petroff, Cleary, DeGeorge, Grunberg, Ali, Appel, Bennett, {et~al.}}]{harrington2018}
Harrington, K., Eimer, J., Chuss, D.~T., {et~al.} 2018, in Millimeter, Submillimeter, and Far-Infrared Detectors and Instrumentation for Astronomy IX, Vol. 10708, SPIE, 611--632

\bibitem[{Hartley \& Zisserman(2004)}]{Hartley2004}
Hartley, R.~I., \& Zisserman, A. 2004, Multiple View Geometry in Computer Vision, 2nd edn. (Cambridge University Press, ISBN: 0521540518)

\bibitem[{Hasselfield {et~al.}(2013)Hasselfield, Moodley, Bond, Das, Devlin, Dunkley, Dünner, Fowler, Gallardo, Gralla, \& et~al.}]{Hasselfield_2013}
Hasselfield, M., Moodley, K., Bond, J.~R., {et~al.} 2013, The Astrophysical Journal Supplement Series, 209, 17, \dodoi{10.1088/0067-0049/209/1/17}

\bibitem[{Hazumi {et~al.}(2020)Hazumi, Ade, Adler, Allys, Arnold, Auguste, Aumont, Aurlien, Austermann, Baccigalupi, \& et~al.}]{Litebird_mission_2020}
Hazumi, M., Ade, P.~A., Adler, A., {et~al.} 2020, Space Telescopes and Instrumentation 2020: Optical, Infrared, and Millimeter Wave, \dodoi{10.1117/12.2563050}

\bibitem[{Hu {et~al.}(2003)Hu, Hedman, \& Zaldarriaga}]{Hu_2003}
Hu, W., Hedman, M.~M., \& Zaldarriaga, M. 2003, Physical Review D, 67, \dodoi{10.1103/physrevd.67.043004}

\bibitem[{{Johnson} {et~al.}(2015){Johnson}, {Vourch}, {Drysdale}, {Kalman}, {Fujikawa}, {Keating}, \& {Kaufman}}]{Johnson2015}
{Johnson}, B.~R., {Vourch}, C.~J., {Drysdale}, T.~D., {et~al.} 2015, Journal of Astronomical Instrumentation, 4, 1550007, \dodoi{10.1142/S2251171715500075}

\bibitem[{{Jost} {et~al.}(2023){Jost}, {Errard}, \& {Stompor}}]{Jost2023}
{Jost}, B., {Errard}, J., \& {Stompor}, R. 2023, \prd, 108, 082005, \dodoi{10.1103/PhysRevD.108.082005}

\bibitem[{Kamionkowski {et~al.}(1997)Kamionkowski, Kosowsky, \& Stebbins}]{Kamionkowski_1997}
Kamionkowski, M., Kosowsky, A., \& Stebbins, A. 1997, Physical Review Letters, 78, 2058–2061, \dodoi{10.1103/physrevlett.78.2058}

\bibitem[{Kamionkowski \& Kovetz(2016)}]{Kamionkowski_2016}
Kamionkowski, M., \& Kovetz, E.~D. 2016, Annual Review of Astronomy and Astrophysics, 54, 227–269, \dodoi{10.1146/annurev-astro-081915-023433}

\bibitem[{{Keating} {et~al.}(2013){Keating}, {Shimon}, \& {Yadav}}]{Keating_2013}
{Keating}, B.~G., {Shimon}, M., \& {Yadav}, A. P.~S. 2013, \apjl, 762, L23, \dodoi{10.1088/2041-8205/762/2/L23}

\bibitem[{Kisner {et~al.}(2021)Kisner, Keskitalo, Zonca, Madsen, Savarit, Tomasi, Cheung, Puglisi, Liu, \& Hasselfield}]{TOAST}
Kisner, T., Keskitalo, R., Zonca, A., {et~al.} 2021, hpc4cmb/toast: Update Pybind11, 2.3.14,  Zenodo, \dodoi{10.5281/zenodo.5559597}

\bibitem[{Kosowsky(1996)}]{Kosowsky_1996}
Kosowsky, A. 1996, Annals of Physics, 246, 49–85, \dodoi{10.1006/aphy.1996.0020}

\bibitem[{Krachmalnicoff {et~al.}(2022)Krachmalnicoff, Matsumura, de~la Hoz, Basak, Gruppuso, Minami, Baccigalupi, Komatsu, Martínez-González, Vielva, Aumont, Aurlien, Azzoni, Banday, Barreiro, Bartolo, Bersanelli, Calabrese, Carones, Casas, Cheung, Chinone, Columbro, de~Bernardis, Diego-Palazuelos, Errard, Finelli, Fuskeland, Galloway, Genova-Santos, Gerbino, Ghigna, Giardiello, Gjerløw, Hazumi, Henrot-Versillé, Kisner, Lamagna, Lattanzi, Levrier, Luzzi, Maino, Masi, Migliaccio, Montier, Morgante, Mot, Nagata, Nati, Natoli, Pagano, Paiella, Paoletti, Patanchon, Piacentini, Polenta, Poletti, Puglisi, Remazeilles, Rubino-Martin, Sasaki, Shiraishi, Signorelli, Stever, Tartari, Tristram, Tsuji, Vacher, Wehus, \& Zannoni}]{Krachmalnicoff_inflight_polang}
Krachmalnicoff, N., Matsumura, T., de~la Hoz, E., {et~al.} 2022, Journal of Cosmology and Astroparticle Physics, 2022, 039, \dodoi{10.1088/1475-7516/2022/01/039}

\bibitem[{Louis {et~al.}(2017)Louis, Grace, Hasselfield, Lungu, Maurin, Addison, Ade, Aiola, Allison, Amiri, Angile, Battaglia, Beall, de~Bernardis, Bond, Britton, Calabrese, Cho, Choi, Coughlin, Crichton, Crowley, Datta, Devlin, Dicker, Dunkley, Dünner, Ferraro, Fox, Gallardo, Gralla, Halpern, Henderson, Hill, Hilton, Hilton, Hincks, Hlozek, Ho, Huang, Hubmayr, Huffenberger, Hughes, Infante, Irwin, Kasanda, Klein, Koopman, Kosowsky, Li, Madhavacheril, Marriage, McMahon, Menanteau, Moodley, Munson, Naess, Nati, Newburgh, Nibarger, Niemack, Nolta, Nuñez, Page, Pappas, Partridge, Rojas, Schaan, Schmitt, Sehgal, Sherwin, Sievers, Simon, Spergel, Staggs, Switzer, Thornton, Trac, Treu, Tucker, Engelen, Ward, \& Wollack}]{Louis_2017}
Louis, T., Grace, E., Hasselfield, M., {et~al.} 2017, Journal of Cosmology and Astroparticle Physics, 2017, 031–031, \dodoi{10.1088/1475-7516/2017/06/031}

\bibitem[{Ludwig(1973)}]{Ludwig1973}
Ludwig, A. 1973, IEEE Transactions on Antennas and Propagation, 21, 116, \dodoi{10.1109/TAP.1973.1140406}

\bibitem[{Lungu {et~al.}(2022)Lungu, Storer, Hasselfield, Duivenvoorden, Calabrese, Chesmore, Choi, Dunkley, Dünner, Gallardo, Golec, Guan, Hill, Hincks, Hubmayr, Madhavacheril, Mallaby-Kay, McMahon, Moodley, Naess, Nati, Niemack, Page, Partridge, Puddu, Schillaci, Sif{\'{o}}n, Staggs, Sunder, Wollack, \& Xu}]{Lungu_2022}
Lungu, M., Storer, E.~R., Hasselfield, M., {et~al.} 2022, Journal of Cosmology and Astroparticle Physics, 2022, 044, \dodoi{10.1088/1475-7516/2022/05/044}

\bibitem[{Manandhar {et~al.}(1999)Manandhar, Honda, \& Murai}]{Manandhar1999}
Manandhar, D., Honda, K., \& Murai, S. 1999, in IEEE 1999 International Geoscience and Remote Sensing Symposium. IGARSS'99 (Cat. No.99CH36293), Vol.~2, 882--884 vol.2, \dodoi{10.1109/IGARSS.1999.774473}

\bibitem[{{McCarrick} {et~al.}(2021){McCarrick}, {Arnold}, {Atkins}, {Austermann}, {Bhandarkar}, {Choi}, {Duell}, {Duff}, {Dutcher}, {Galitzk}, {Healy}, {Huber}, {Hubmayr}, {Johnson}, {Niemack}, {Seibert}, {Silva-Feaver}, {Sonka}, {Staggs}, {Vavagiakis}, {Wang}, {Xu}, {Zheng}, \& {Zhu}}]{McCarrick2021}
{McCarrick}, H., {Arnold}, K., {Atkins}, Z., {et~al.} 2021, arXiv e-prints, arXiv:2112.01458, \dodoi{10.48550/arXiv.2112.01458}

\bibitem[{Miller {et~al.}(2009)Miller, Shimon, \& Keating}]{Miller_beam_asymmetry}
Miller, N.~J., Shimon, M., \& Keating, B.~G. 2009, Physical Review D, 79, \dodoi{10.1103/physrevd.79.103002}

\bibitem[{{Minami} {et~al.}(2019){Minami}, {Ochi}, {Ichiki}, {Katayama}, {Komatsu}, \& {Matsumura}}]{Minami_2019}
{Minami}, Y., {Ochi}, H., {Ichiki}, K., {et~al.} 2019, arXiv e-prints, arXiv:1904.12440, \dodoi{10.48550/arXiv.1904.12440}

\bibitem[{Murai {et~al.}(2023)Murai, Naokawa, Namikawa, \& Komatsu}]{cosmic_birefringence}
Murai, K., Naokawa, F., Namikawa, T., \& Komatsu, E. 2023, Phys. Rev. D, 107, L041302, \dodoi{10.1103/PhysRevD.107.L041302}

\bibitem[{{Murphy} {et~al.}(2024){Murphy}, {Choi}, {Datta}, {Devlin}, {Hasselfield}, {Koopman}, {McMahon}, {Naess}, {Niemack}, {Page}, {Staggs}, {Thornton}, \& {Wollack}}]{Murphy_2024}
{Murphy}, C.~C., {Choi}, S.~K., {Datta}, R., {et~al.} 2024, \ao, 63, 5079, \dodoi{10.1364/AO.521079}

\bibitem[{Nati {et~al.}(2017)Nati, Devlin, Gerbino, Johnson, Keating, Pagano, \& Teply}]{Nati2017}
Nati, F., Devlin, M.~J., Gerbino, M., {et~al.} 2017, Journal of Astronomical Instrumentation, 06, 1740008, \dodoi{10.1142/S2251171717400086}

\bibitem[{{Navaroli} {et~al.}(2018){Navaroli}, {Teply}, {Crowley}, {Kaufman}, {Galitzki}, {Arnold}, \& {Keating}}]{Navaroli_2018}
{Navaroli}, M.~F., {Teply}, G.~P., {Crowley}, K.~D., {et~al.} 2018, in Society of Photo-Optical Instrumentation Engineers (SPIE) Conference Series, Vol. 10708, Millimeter, Submillimeter, and Far-Infrared Detectors and Instrumentation for Astronomy IX, ed. J.~{Zmuidzinas} \& J.-R. {Gao}, 107082A, \dodoi{10.1117/12.2312856}

\bibitem[{on~Illumination(2004)}]{international2004colorimetry}
on~Illumination, I.~C. 2004, Colorimetry, CIE technical report ; 15:2004 (Commission Internationale de l'Eclairage).
\newblock \url{https://books.google.it/books?id=P1NkAAAACAAJ}

\bibitem[{{Planck Collaboration} {et~al.}(2017){Planck Collaboration}, Akrami, Ashdown, Aumont, Baccigalupi, Ballardini, Banday, Barreiro, Bartolo, Basak, Benabed, Bernard, Bersanelli, Bielewicz, Bonavera, Bond, Borrill, Bouchet, Boulanger, Bucher, Burigana, Butler, Calabrese, Cardoso, Carron, Chiang, Colombo, Comis, Couchot, Coulais, Crill, Curto, Cuttaia, de~Bernardis, de~Rosa, de~Zotti, Delabrouille, Di~Valentino, Dickinson, Diego, Doré, Ducout, Dupac, Elsner, Enßlin, Eriksen, Falgarone, Fantaye, Finelli, Frailis, Fraisse, Franceschi, Frolov, Galeotta, Galli, Ganga, Génova-Santos, Gerbino, González-Nuevo, Górski, Gruppuso, Gudmundsson, Hansen, Helou, Henrot-Versillé, Herranz, Hivon, Jaffe, Jones, Keihänen, Keskitalo, Kiiveri, Kim, Kisner, Krachmalnicoff, Kunz, Kurki-Suonio, Lagache, Lamarre, Lasenby, Lattanzi, Lawrence, Le~Jeune, Lellouch, Levrier, Liguori, Lilje, Lindholm, López-Caniego, Ma, Macías-Pérez, Maggio, Maino, Mandolesi, Maris, Martin, Martínez-González, Matarrese, Mauri, McEwen,
  Melchiorri, Mennella, Migliaccio, Miville-Deschênes, Molinari, Moneti, Montier, Moreno, Morgante, Natoli, Oxborrow, Paoletti, Partridge, Patanchon, Patrizii, Perdereau, Piacentini, Plaszczynski, Polenta, Rachen, Racine, Reinecke, Remazeilles, Renzi, Rocha, Romelli, Rosset, Roudier, Rubiño-Martín, Ruiz-Granados, Salvati, Sandri, Savelainen, Scott, Sirri, Spencer, Suur-Uski, Tauber, Tavagnacco, Tenti, Toffolatti, Tomasi, Tristram, Trombetti, Valiviita, Van~Tent, Vielva, Villa, Wehus, \& Zacchei}]{Planck_LII_2017}
{Planck Collaboration}, Akrami, Y., Ashdown, M., {et~al.} 2017, \aap, 607, A122, \dodoi{10.1051/0004-6361/201630311}

\bibitem[{{Planck Collaboration} {et~al.}(2020){Planck Collaboration}, {Akrami}, {Arroja}, {Ashdown}, {Aumont}, {Baccigalupi}, {Ballardini}, {Banday}, {Barreiro}, {Bartolo}, {Basak}, {Benabed}, {Bernard}, {Bersanelli}, {Bielewicz}, {Bond}, {Borrill}, {Bouchet}, {Bucher}, {Burigana}, {Butler}, {Calabrese}, {Cardoso}, {Casaponsa}, {Challinor}, {Chiang}, {Colombo}, {Combet}, {Crill}, {Cuttaia}, {de Bernardis}, {de Rosa}, {de Zotti}, {Delabrouille}, {Delouis}, {Di Valentino}, {Diego}, {Dor{\'e}}, {Douspis}, {Ducout}, {Dupac}, {Dusini}, {Efstathiou}, {Elsner}, {En{\ss}lin}, {Eriksen}, {Fantaye}, {Fergusson}, {Fernandez-Cobos}, {Finelli}, {Frailis}, {Fraisse}, {Franceschi}, {Frolov}, {Galeotta}, {Galli}, {Ganga}, {G{\'e}nova-Santos}, {Gerbino}, {Gonz{\'a}lez-Nuevo}, {G{\'o}rski}, {Gratton}, {Gruppuso}, {Gudmundsson}, {Hamann}, {Handley}, {Hansen}, {Herranz}, {Hivon}, {Huang}, {Jaffe}, {Jones}, {Jung}, {Keih{\"a}nen}, {Keskitalo}, {Kiiveri}, {Kim}, {Krachmalnicoff}, {Kunz}, {Kurki-Suonio}, {Lamarre}, {Lasenby},
  {Lattanzi}, {Lawrence}, {Le Jeune}, {Levrier}, {Lewis}, {Liguori}, {Lilje}, {Lindholm}, {L{\'o}pez-Caniego}, {Ma}, {Mac{\'\i}as-P{\'e}rez}, {Maggio}, {Maino}, {Mandolesi}, {Marcos-Caballero}, {Maris}, {Martin}, {Mart{\'\i}nez-Gonz{\'a}lez}, {Matarrese}, {Mauri}, {McEwen}, {Meerburg}, {Meinhold}, {Melchiorri}, {Mennella}, {Migliaccio}, {Miville-Desch{\^e}nes}, {Molinari}, {Moneti}, {Montier}, {Morgante}, {Moss}, {M{\"u}nchmeyer}, {Natoli}, {Oppizzi}, {Pagano}, {Paoletti}, {Partridge}, {Patanchon}, {Perrotta}, {Pettorino}, {Piacentini}, {Polenta}, {Puget}, {Rachen}, {Racine}, {Reinecke}, {Remazeilles}, {Renzi}, {Rocha}, {Rubi{\~n}o-Mart{\'\i}n}, {Ruiz-Granados}, {Salvati}, {Savelainen}, {Scott}, {Shellard}, {Shiraishi}, {Sirignano}, {Sirri}, {Smith}, {Spencer}, {Stanco}, {Sunyaev}, {Suur-Uski}, {Tauber}, {Tavagnacco}, {Tenti}, {Toffolatti}, {Tomasi}, {Trombetti}, {Valiviita}, {Van Tent}, {Vielva}, {Villa}, {Vittorio}, {Wandelt}, {Wehus}, {Zacchei}, \& {Zonca}}]{Planck_IX_2020}
{Planck Collaboration}, {Akrami}, Y., {Arroja}, F., {et~al.} 2020, \aap, 641, A9, \dodoi{10.1051/0004-6361/201935891}

\bibitem[{{Planck Collaboration V}(2020)}]{Planck_V_2020}
{Planck Collaboration V}. 2020, \aap, 641, A5, \dodoi{10.1051/0004-6361/201936386}

\bibitem[{{Planck Collaboration VI}(2020)}]{Planck_VI_2020}
{Planck Collaboration VI}. 2020, \aap, 641, A6, \dodoi{10.1051/0004-6361/201833910}

\bibitem[{Seljak \& Zaldarriaga(1997)}]{Seljak_1997}
Seljak, U., \& Zaldarriaga, M. 1997, Physical Review Letters, 78, 2054–2057, \dodoi{10.1103/physrevlett.78.2054}

\bibitem[{Sharma {et~al.}(2005)Sharma, Wu, \& Dalal}]{Sharma2005}
Sharma, G., Wu, W., \& Dalal, E.~N. 2005, Color Research \& Application, 30, 21, \dodoi{10.1002/col.20070}

\bibitem[{{The BICEP Collaboration III}(2015)}]{bicep2_systematics}
{The BICEP Collaboration III}. 2015, The Astrophysical Journal, 814, 110, \dodoi{10.1088/0004-637x/814/2/110}

\bibitem[{{The Planck Collaboration VII} {et~al.}(2016){The Planck Collaboration VII}, Adam, Ade, Aghanim, Arnaud, Ashdown, Aumont, Baccigalupi, Banday, Barreiro, Bartolo, Battaner, Benabed, Benoît, Benoit-Lévy, Bernard, Bersanelli, Bertincourt, Bielewicz, Bock, Bonavera, Bond, Borrill, Bouchet, Boulanger, Bucher, Burigana, Calabrese, Cardoso, Catalano, Challinor, Chamballu, Chary, Chiang, Christensen, Clements, Colombi, Colombo, Combet, Couchot, Coulais, Crill, Curto, Cuttaia, Danese, Davies, Davis, de~Bernardis, de~Rosa, de~Zotti, Delabrouille, Delouis, Désert, Diego, Dole, Donzelli, Doré, Douspis, Ducout, Dupac, Efstathiou, Elsner, Enßlin, Eriksen, Falgarone, Fergusson, Finelli, Forni, Frailis, Fraisse, Franceschi, Frejsel, Galeotta, Galli, Ganga, Ghosh, Giard, Giraud-Héraud, Gjerløw, González-Nuevo, Górski, Gratton, Gruppuso, Gudmundsson, Hansen, Hanson, Harrison, Henrot-Versillé, Herranz, Hildebrandt, Hivon, Hobson, Holmes, Hornstrup, Hovest, Huffenberger, Hurier, Jaffe, Jaffe, Jones, Juvela,
  Keihänen, Keskitalo, Kisner, Kneissl, Knoche, Kunz, Kurki-Suonio, Lagache, Lamarre, Lasenby, Lattanzi, Lawrence, Le~Jeune, Leahy, Lellouch, Leonardi, Lesgourgues, Levrier, Liguori, Lilje, Linden-Vørnle, López-Caniego, Lubin, Macías-Pérez, Maggio, Maino, Mandolesi, Mangilli, Maris, Martin, Martínez-González, Masi, Matarrese, McGehee, Melchiorri, Mendes, Mennella, Migliaccio, Mitra, Miville-Deschênes, Moneti, Montier, Moreno, Morgante, Mortlock, Moss, Mottet, Munshi, Murphy, Naselsky, Nati, Natoli, Netterfield, Nørgaard-Nielsen, Noviello, Novikov, Novikov, Oxborrow, Paci, Pagano, Pajot, Paoletti, Pasian, Patanchon, Pearson, Perdereau, Perotto, Perrotta, Pettorino, Piacentini, Piat, Pierpaoli, Pietrobon, Plaszczynski, Pointecouteau, Polenta, Pratt, Prézeau, Prunet, Puget, Rachen, Reinecke, Remazeilles, Renault, Renzi, Ristorcelli, Rocha, Rosset, Rossetti, Roudier, Rowan-Robinson, Rusholme, Sandri, Santos, Sauvé, Savelainen, Savini, Scott, Seiffert, Shellard, Spencer, Stolyarov, Stompor, Sudiwala,
  Sutton, Suur-Uski, Sygnet, Tauber, Terenzi, Toffolatti, Tomasi, Tristram, Tucci, Tuovinen, Valenziano, Valiviita, Van~Tent, Vibert, Vielva, Villa, Wade, Wandelt, Watson, Wehus, Yvon, Zacchei, \& Zonca}]{planck_vii_2016}
{The Planck Collaboration VII}, Adam, R., Ade, P. A.~R., {et~al.} 2016, \aap, 594, A7, \dodoi{10.1051/0004-6361/201525844}

\bibitem[{{The Simons Observatory} {et~al.}(2019){The Simons Observatory}, Ade, Aguirre, Ahmed, Aiola, Ali, Alonso, Alvarez, Arnold, Ashton, Austermann, Awan, Baccigalupi, Baildon, Barron, Battaglia, Battye, Baxter, Bazarko, Beall, Bean, Beck, Beckman, Beringue, Bianchini, Boada, Boettger, Bond, Borrill, Brown, Bruno, Bryan, Calabrese, Calafut, Calisse, Carron, Challinor, Chesmore, Chinone, Chluba, Cho, Choi, Coppi, Cothard, Coughlin, Crichton, Crowley, Crowley, Cukierman, D’Ewart, Dünner, de~Haan, Devlin, Dicker, Didier, Dobbs, Dober, Duell, Duff, Duivenvoorden, Dunkley, Dusatko, Errard, Fabbian, Feeney, Ferraro, Fluxà, Freese, Frisch, Frolov, Fuller, Fuzia, Galitzki, Gallardo, Ghersi, Gao, Gawiser, Gerbino, Gluscevic, Goeckner-Wald, Golec, Gordon, Gralla, Green, Grigorian, Groh, Groppi, Guan, Gudmundsson, Han, Hargrave, Hasegawa, Hasselfield, Hattori, Haynes, Hazumi, He, Healy, Henderson, Hervias-Caimapo, Hill, Hill, Hilton, Hilton, Hincks, Hinshaw, Hložek, Ho, Ho, Howe, Huang, Hubmayr, Huffenberger,
  Hughes, Ijjas, Ikape, Irwin, Jaffe, Jain, Jeong, Kaneko, Karpel, Katayama, Keating, Kernasovskiy, Keskitalo, Kisner, Kiuchi, Klein, Knowles, Koopman, Kosowsky, Krachmalnicoff, Kuenstner, Kuo, Kusaka, Lashner, Lee, Lee, Leon, Leung, Lewis, Li, Li, Limon, Linder, Lopez-Caraballo, Louis, Lowry, Lungu, Madhavacheril, Mak, Maldonado, Mani, Mates, Matsuda, Maurin, Mauskopf, May, McCallum, McKenney, McMahon, Meerburg, Meyers, Miller, Mirmelstein, Moodley, Munchmeyer, Munson, Naess, Nati, Navaroli, Newburgh, Nguyen, Niemack, Nishino, Orlowski-Scherer, Page, Partridge, Peloton, Perrotta, Piccirillo, Pisano, Poletti, Puddu, Puglisi, Raum, Reichardt, Remazeilles, Rephaeli, Riechers, Rojas, Roy, Sadeh, Sakurai, Salatino, Rao, Schaan, Schmittfull, Sehgal, Seibert, Seljak, Sherwin, Shimon, Sierra, Sievers, Sikhosana, Silva-Feaver, Simon, Sinclair, Siritanasak, Smith, Smith, Spergel, Staggs, Stein, Stevens, Stompor, Suzuki, Tajima, Takakura, Teply, Thomas, Thorne, Thornton, Trac, Tsai, Tucker, Ullom, Vagnozzi, Engelen,
  Lanen, Winkle, Vavagiakis, Vergès, Vissers, Wagoner, Walker, Ward, Westbrook, Whitehorn, Williams, Williams, Wollack, Xu, Yu, Yu, Zago, Zhang, \& Zhu}]{SO_goals_forecasts}
{The Simons Observatory}, Ade, P., Aguirre, J., {et~al.} 2019, Journal of Cosmology and Astroparticle Physics, 2019, 056–056, \dodoi{10.1088/1475-7516/2019/02/056}

\bibitem[{Tristram {et~al.}(2021)Tristram, Banday, Górski, Keskitalo, Lawrence, Andersen, Barreiro, Borrill, Eriksen, Fernandez-Cobos, Kisner, Martínez-González, Partridge, Scott, Svalheim, Thommesen, \& Wehus}]{Tristram_2021}
Tristram, M., Banday, A.~J., Górski, K.~M., {et~al.} 2021, \aap, 647, A128, \dodoi{10.1051/0004-6361/202039585}

\bibitem[{{Verg{\`e}s} {et~al.}(2021){Verg{\`e}s}, {Errard}, \& {Stompor}}]{Verges2021}
{Verg{\`e}s}, C., {Errard}, J., \& {Stompor}, R. 2021, \prd, 103, 063507, \dodoi{10.1103/PhysRevD.103.063507}

\bibitem[{{Vielva} {et~al.}(2022){Vielva}, {Mart{\'\i}nez-Gonz{\'a}lez}, {Casas}, {Matsumura}, {Henrot-Versill{\'e}}, {Komatsu}, {Aumont}, {Aurlien}, {Baccigalupi}, {Banday}, {Barreiro}, {Bartolo}, {Calabrese}, {Cheung}, {Columbro}, {Coppolecchia}, {de Bernardis}, {de Haan}, {de la Hoz}, {De Petris}, {Della Torre}, {Diego-Palazuelos}, {Eriksen}, {Errard}, {Finelli}, {Franceschet}, {Fuskeland}, {Galloway}, {Ganga}, {Gervasi}, {G{\'e}nova-Santos}, {Ghigna}, {Gjerl{\o}w}, {Gruppuso}, {Hazumi}, {Herranz}, {Hivon}, {Kohri}, {Lamagna}, {Leloup}, {Macias-Perez}, {Masi}, {Matsuda}, {Morgante}, {Nakano}, {Nati}, {Natoli}, {Nerval}, {Odagiri}, {Oguri}, {Pagano}, {Paiella}, {Paoletti}, {Piacentini}, {Polenta}, {Puglisi}, {Remazeilles}, {Ritacco}, {Rubino-Martin}, {Scott}, {Sekimoto}, {Shiraishi}, {Signorelli}, {Takakura}, {Tartari}, {Thompson}, {Tristram}, {Vacher}, {Vittorio}, {Wehus}, {Zannoni}, \& {The LiteBIRD collaboration}}]{Vielva2022}
{Vielva}, P., {Mart{\'\i}nez-Gonz{\'a}lez}, E., {Casas}, F.~J., {et~al.} 2022, \jcap, 2022, 029, \dodoi{10.1088/1475-7516/2022/04/029}

\bibitem[{Weiland {et~al.}(2011)Weiland, Odegard, Hill, Wollack, Hinshaw, Greason, Jarosik, Page, Bennett, Dunkley, \& et~al.}]{calibration_wmap_2011}
Weiland, J.~L., Odegard, N., Hill, R.~S., {et~al.} 2011, The Astrophysical Journal Supplement Series, 192, 19, \dodoi{10.1088/0067-0049/192/2/19}

\bibitem[{Yamada {et~al.}(2024)Yamada, Bixler, Sakurai, Ashton, Sugiyama, Arnold, Begin, Corbett, Day-Weiss, Galitzki, {et~al.}}]{yamada2024simons}
Yamada, K., Bixler, B., Sakurai, Y., {et~al.} 2024, Review of Scientific Instruments, 95, \dodoi{10.1063/5.0252360}

\end{thebibliography}
\bibliographystyle{aasjournal}

\end{document}